\documentclass[a4paper,11pt]{article}
\pdfoutput=1

\usepackage{jheppub}

\usepackage[utf8]{inputenc}

\usepackage{bbold}
\usepackage{subfigure}
\usepackage{epsfig}
\usepackage[dvipsnames]{xcolor}
\usepackage{amssymb}
\usepackage{multicol}
\def\simle{\mathrel{\rlap{\raise 0.511ex \hbox{$<$}}{\lower 0.511ex \hbox{$\sim$}}}}
\newcommand{\bea}{\begin{eqnarray}}
\newcommand{\beq}{\begin{equation}}
\newcommand{\eea}{\end{eqnarray}}
\newcommand{\eeq}{\end{equation}}

\newcommand{\vep}{\varepsilon}

\newcommand{\nn}{\nonumber}

\newcommand{\lsim}{\raise0.3ex\hbox{$\;<$\kern-0.75em\raise-1.1ex\hbox{$\sim\;$}}}
\newcommand{\gsim}{\raise0.3ex\hbox{$\;>$\kern-0.75em\raise-1.1ex\hbox{$\sim\;$}}}
\newcommand{\eq}[1]{Eq.~(\ref{#1})}
\newcommand{\unity}{{\hbox{1\kern-.8mm l}}}

\title{\boldmath Controlled flavor violation in the MSSM from a unified $\Delta(27)$ flavor symmetry}

\author[a]{Ivo de Medeiros Varzielas}
\author[b,\, c]{M.L. L\'opez-Ib\'a\~nez}
\author[b]{Aurora Melis,}
\author[b]{and Oscar Vives}

\affiliation[a]{CFTP, Departamento de F\'isica, Instituto Superior T\'ecnico, Universidade de Lisboa\\ Avenida Rovisco Pais 1, 1049 Lisboa, Portugal}
\affiliation[b]{Departament de F\'{i}sica T\`{e}orica, Universitat de Val\`{e}ncia and IFIC, Universitat de Val\`{e}ncia-CSIC  \\
	Dr. Moliner 50, E-46100 Burjassot (Val\`{e}ncia), Spain}
\affiliation[c]{Dip. di Matematica e Fisica, Università di Roma Tre and
INFN, Sezione di Roma III \\
Via della Vasca Navale 84, 00146 Rome, Italy}    

\emailAdd{ivo.de@udo.edu}
\emailAdd{maloi2@uv.es}
\emailAdd{aurora.melis@ific.uv.es}
\emailAdd{oscar.vives@uv.es}

\preprint{FTUV-18-0628, IFIC-18-26}

\abstract{
We study the phenomenology of a unified supersymmetric theory with a flavor symmetry $\Delta(27)$. The model accommodates quark and lepton masses, mixing angles and CP phases. In this model, the Dirac and Majorana mass matrices have a unified texture zero structure in the $(1,1)$ entry that leads to the Gatto-Sartori-Tonin relation between the Cabibbo angle and ratios of the masses in the quark sectors, and to a natural departure from zero of the $\theta_{13}^\ell$ angle in the lepton sector. We derive the flavor structures of the trilinears and soft mass matrices, and show their general non-universality. This causes large flavor violating effects. As a consequence, the parameter space for this model is constrained, allowing it to be (dis)proven by flavor violation searches in the next decade. Although the results are model specific, we compare them to previous studies to show similar flavour effects (and associated constraints) are expected in general in supersymmetric flavor models, and may be used to distinguish them.
}

\begin{document} 
\maketitle
\flushbottom
\setlength{\parindent}{0in}

\section{Introduction}
\label{sec:intro}

Using symmetries to interpret the chaotic picture of flavor parameters in the SM is a well-known and developed strategy. Nevertheless, a univocal picture has not emerged, driving to a plethora of viable choices for the flavor symmetry $\mathcal{G}_f$ and for its breaking, which are consistent with the observed fermionic masses and mixig angles. Likely, the only possibility to disentangle the puzzle of the origin of flavor is to discover flavor-sensitive New Physics (NP). Supersymmetric extensions of the Standard Model (SM) give a good example in this sense, where in addition to the usual Yukawa couplings of the SM we have the soft breaking terms: the trilinears and soft-mass matrices corresponding to the scalar superpartners. Under the requirement that the mediation of Supersymmetry breaking to the visible sector occurs at a higher scale $\Lambda_{\rm Med}$ than the breaking of the flavor symmetry $\Lambda_f$, i.e. $\Lambda_{\rm Med} \gg\Lambda_f$, these three flavor structures will have to respect the same $\mathcal{G}_f$ and, after the breaking of the symmetry, be similarly non-trivial.

In \cite{Das:2016czs,Lopez-Ibanez:2017xxw} we investigated the case where Supersymmetry breaking is communicated through a spurion field, $X$, coupling universally to the visible sector, and showed that we can expect a mismatch between the Yukawa, Kinetic matrices, and the soft-breaking terms, that prevents the simultaneous diagonalization of the four structures. The mismatch is simply given by the different ways in which the $X$-field may be inserted in the full theory diagrams and its calculation is straightforward in terms of the operator dimension of the terms entering in the Superpotential and K\"{a}hler potential.

In this work, we have applied this type of analysis to a recent model with a unified texture zero structure in the $(1,1)$ entry \cite{deMedeirosVarzielas:2017sdv}. This is an appealing flavor model as it is consistent with an underlying $SO(10)$ grand unification and makes several important postdictions, for example the Gatto-Sartori-Tonin relation between the Cabibbo angle and the quark mass ratios:
\begin{equation}
\sin{\theta _c} = \left\vert \sqrt {\frac{{{m_d}}}{{{m_s}}}}  - {e^{i\delta }}\sqrt {\frac{{{m_u}}}{{{m_c}}}} \right\vert
\label{GST}
\end{equation}
Additionally it predicts the phenomenologically successful trimaximal 1 mixing scheme for the leptons \cite{Albright:2010ap,Varzielas:2012pa}.
Nevertheless, it is important to find additional ways to constrain this and other flavor models, in order to better distinguish between models which, by necessity of the experimentally observed values, make similar postdictions for the fermion masses and mixing angles. Flavor violating (FV) effects associated with new particles and interactions provide one of the best options for constraining flavor models, and this applies in particular to supersymmetric flavor models.

The layout of the paper is as follows. In Section \ref{sec:mech}, a summary of the main results of the mechanism is provided. In Section \ref{sec:model}, we review some relevant details about the model. In Section \ref{sec:analysis}, we present the analysis of FV processes, showing the exclusion regions that constrain the parameter space of the model. We conclude in Section \ref{sec:conclusion}.

\section{A review of the mechanism \label{sec:mech}}

Here we outline the main results of previous works \cite{Das:2016czs,Lopez-Ibanez:2017xxw}, showing that strongly non-universal structures generally arise in SUSY models augmented with a flavor symmetry broken at a scale $\Lambda_f \ll \Lambda_{\rm Med}$. This is true even in the case in which the full theory is totally flavor blind at high energies, e.g. if the breaking of supersymmetry is parametrized by a single field, spurion $X$, universally coupled to the visible sector.

An example of this type of models is supergravity, depending only on the traditional supergravity input parameters $\{m_0\,,\,M_{1/2}\,,\,a_0\,,\,\tan\beta\,,\,|\mu|\}$. In its simplest form, it gives rise to the Minimal Supersymmetric Standard Model (MSSM), completely defined by the usual particle content, superpotential interactions and soft-breaking terms. In this scenario, supersymmetry breaking in a hidden sector is caused by a non-vanishing F-term, $\langle X\rangle = F_{X}$, and propagated to the visible sector through gravitational interactions, suppressed by the Planck scale $M_{\rm Pl}$. In this way, operators analogous to the ones generating the Yukawa couplings with an additional $X$ field enter the Superpotential (${\cal W}_{\Psi}$) and Kähler potential (${\cal K}_{\Psi}$), generating the trilinears and soft-breaking masses:

\beq \label{eqn:Lsoft}
{\cal L}_{\rm soft} ~ = ~ \frac{F_{X}}{ M_{\rm Pl}}\: {\cal W}_{\Psi} \;+\; \frac{F_{X}\: F_{X}^\dagger}{ M_{\rm Pl}^2}\; {\cal K}_{\Psi}\,,
\eeq

where $\Psi$ represents any of the MSSM fields in the visible sector.
Introducing a family symmetry in this context means that all these structures must be symmetric under the group transformations above $\Lambda_f$.  
Initially, the standard Yukawas, with the possible exception of the top-quark Yukawa coupling, are forbidden by the symmetry and they only appear after spontanous breaking of flavor, as powers of an expansion parameter, {\small $\vep \equiv \langle {\Phi}\rangle/M \ll 1$}. Similarly, the Kinetic terms and the soft-breaking interactions will receive corrections from non-renormalizable operators. Then, the superpotential and Kähler potential may be expressed as:

\bea
{\cal W}_{\Psi}  &=& {\cal W}_{\Psi}^{\rm (ren)} \:+\: {\Psi\, \bar\Psi\, H}\; \sum_{\Phi}\sum_{n_{\rm in}=1}^{\infty}\, x_{{\rm n_{in}}}\, \left(\frac{ \langle\Phi\rangle}{\rm M}\right)^{n_{\rm in}} \hspace{5mm}\label{eqn:superpot}
\nn\\[2pt]
{\cal K}_{\Psi}  &=&  {\Psi}\, {\Psi}^\dagger\: \left[\, \mathbb{1}\:+\: \sum_{\Phi,\Phi^\dagger}\sum_{\substack{n_{\rm in},\\ \quad n_{\rm out}=1}}^{\infty}\, c_{({\rm n_{in},n_{out} })}\, \left(\frac{ \langle\Phi\rangle}{\rm M}\right)^{n_{\rm in}}\left(\frac{ \langle\Phi^\dagger\rangle}{\rm M}\right)^{n_{\rm out}}\,\right] \,,\label{eqn:kahler}\\
\nn
\eea
where we are summing over all the flavons in the model and the number of their possible insertions $n_{\rm in}$ (or $n_{\rm out}$, in the case of daggered fields). \\
Tree level FV-effects arise at low energies as a consequence of the misalignment and non-universality between the supersymmetric and the soft-breaking terms when integrating out flavor mediators. Let us see this in detail starting with the superpotential.

\begin{figure}[t!] 
\center
\includegraphics[width=0.45\textwidth]{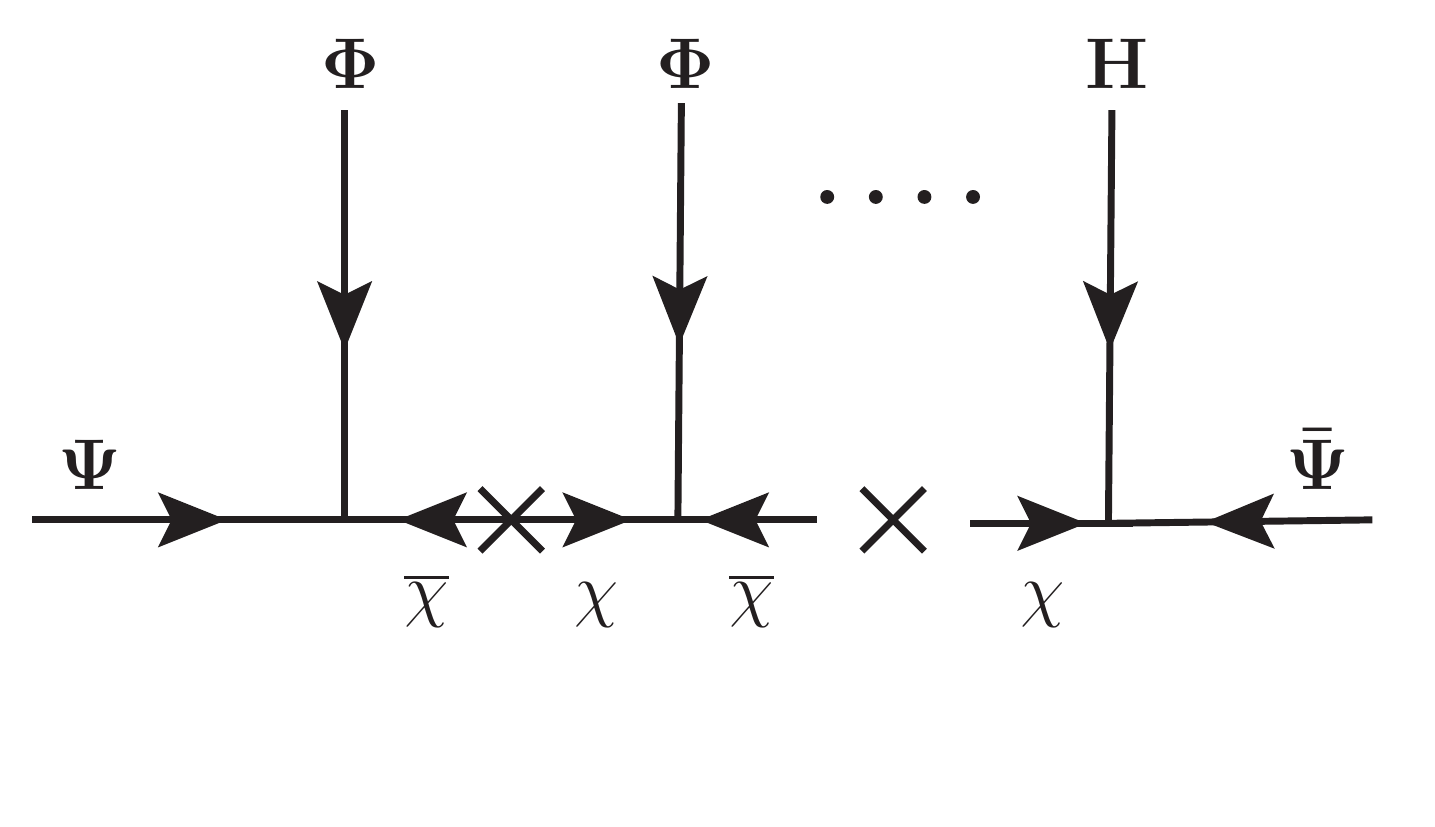}  
\caption{A supergraph depiction of the corrections to the superpotential represented by Eq.~\ref{eqn:superpot}. The internal lines are heavy messengers, and the cross denotes a supersymmetric mass insertion $\rm M$.  
} \label{fig:superpot}
\end{figure}

\begin{figure}[t!]
\center
\includegraphics[width=1.\textwidth]{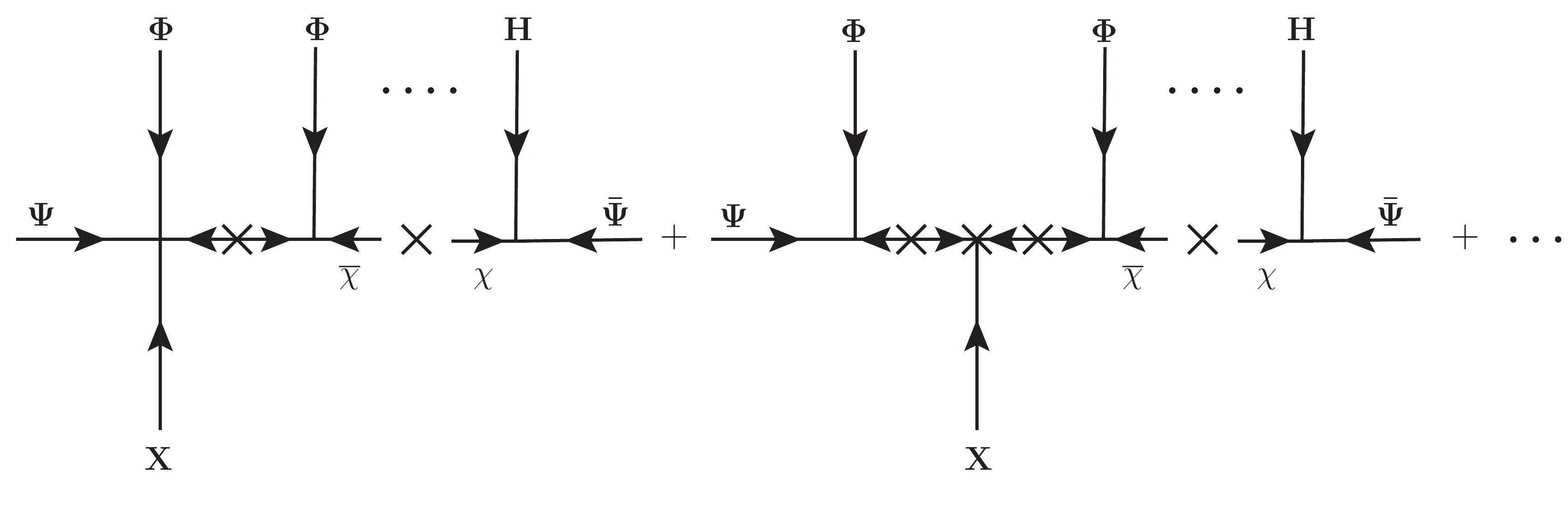} \\[-0.3cm]      
\caption{A supergraph representation of the corrections to the Trilinear couplings, the first term in Eq.~\ref{eqn:Lsoft}. Since the spurion field, $X$, can be attached at any of the vertices, ($2\,n_{\rm in}+1$) diagrams contribute to the same effective operator.
} \label{fig:trilinear}
\end{figure}

A typical diagram responsible for the first term in \eq{eqn:superpot} is depicted in Figure \ref{fig:superpot}, where the internal lines represent heavy messengers integrated out in the effective theory. From here, and still using this pictorial representation, the corresponding $A$-term (first element in \eq{eqn:Lsoft}) may be computed by simply attaching one $X$-insertion in the previous diagram. However, it should be noted that there are multiple ways in which this can be done: $X$ can be inserted at any of the vertices and all these graphs will contribute to the same effective trilinear term, see Figure \ref{fig:trilinear}. Thus, for each Yukawa element $Y_{ij}$ the corresponding Trilinear $A_{ij}$ will be:

\beq \label{eqn:factors1}
A_{ij} \:\propto\: (2\,n_{\rm in}+1)\; a_0\; Y_{ij},
\eeq

with {$a_0\equiv k\, m_0$}, $k\sim \mathcal{O}(1)$ and { $m_0\equiv\langle F_X \rangle/M_{\rm P}$}. Taking into account that, in flavor models, each entry in the Yukawa matrix involves a different number of flavon insertions, the total Trilinear matrix will not be directly proportional to the Yukawa matrix and therefore the rotation to the mass basis (diagonal Yukawas) will not diagonalize the $A$-terms in general. Effects on flavor violating observables will become visible due to the surviving off-diagonal elements.

\begin{figure}[h!]
\center
\subfigure{\includegraphics[width=0.5\textwidth]
{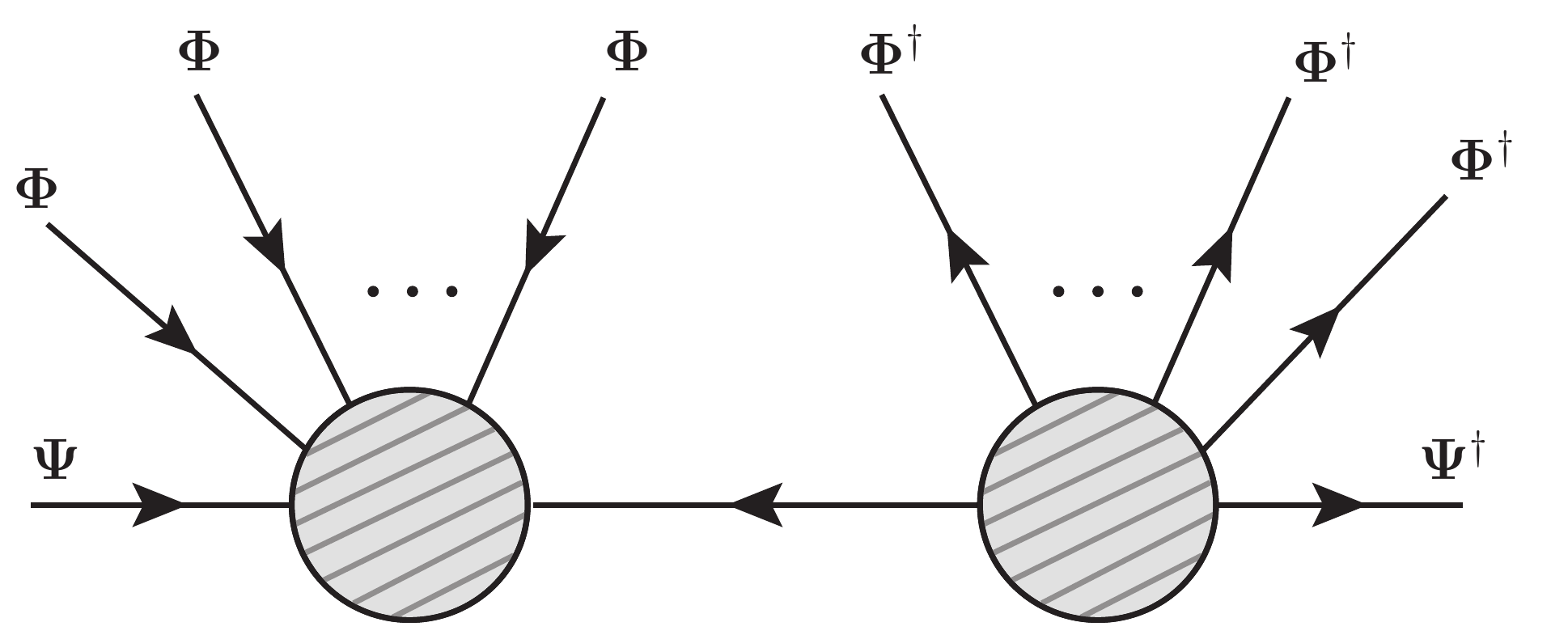}}    
\subfigure{\includegraphics[width=0.66\textwidth]{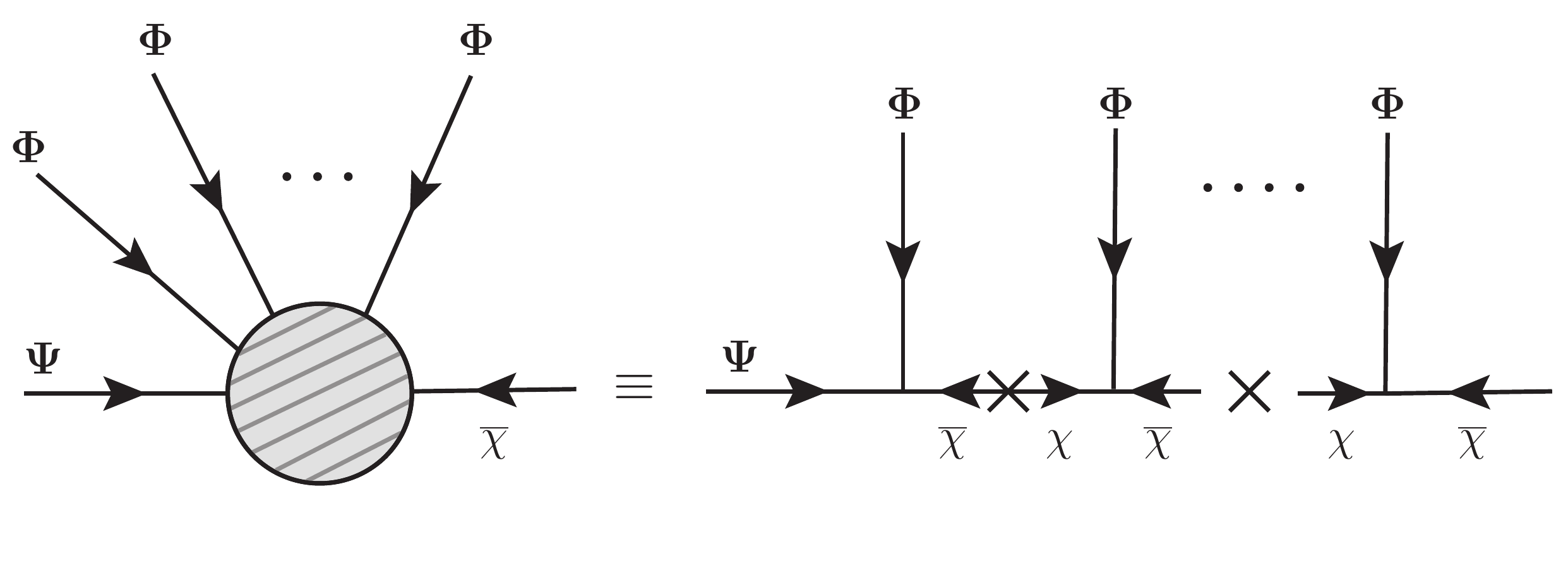} }
\caption{A supergraph depiction of the corrections to the K\"{a}hler potential in Eq.~\ref{eqn:kahler}.\label{fig:kahler}}
\end{figure}

\begin{figure}[h!] 
	\begin{minipage}[r]{0.5\textwidth}
	 	\subfigure[]{\includegraphics[width=0.9\textwidth]{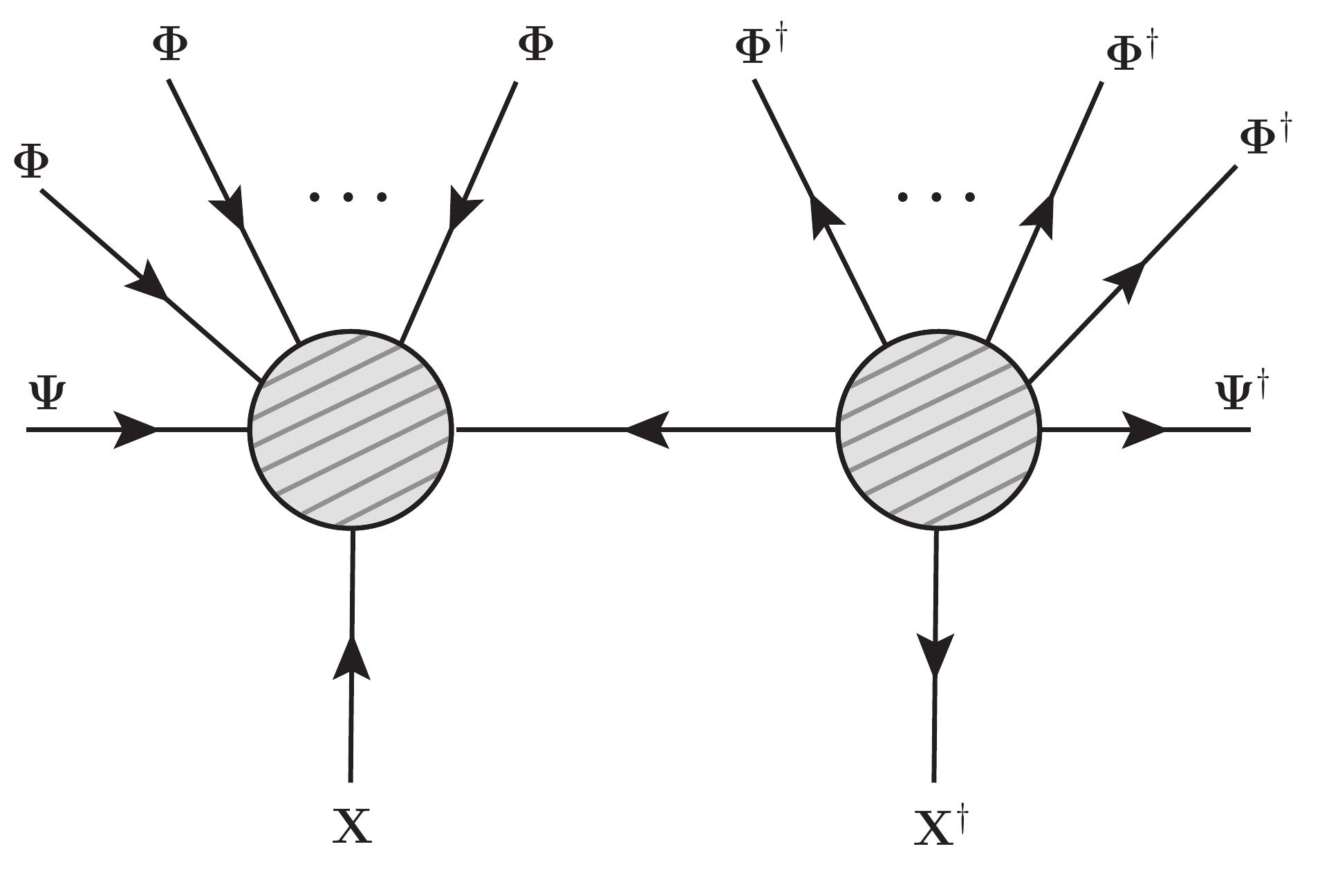}}
    \end{minipage}
    \hspace{0.7cm}
    \begin{minipage}[l]{0.5\textwidth}
		\subfigure[]{\includegraphics[width=0.9\textwidth]{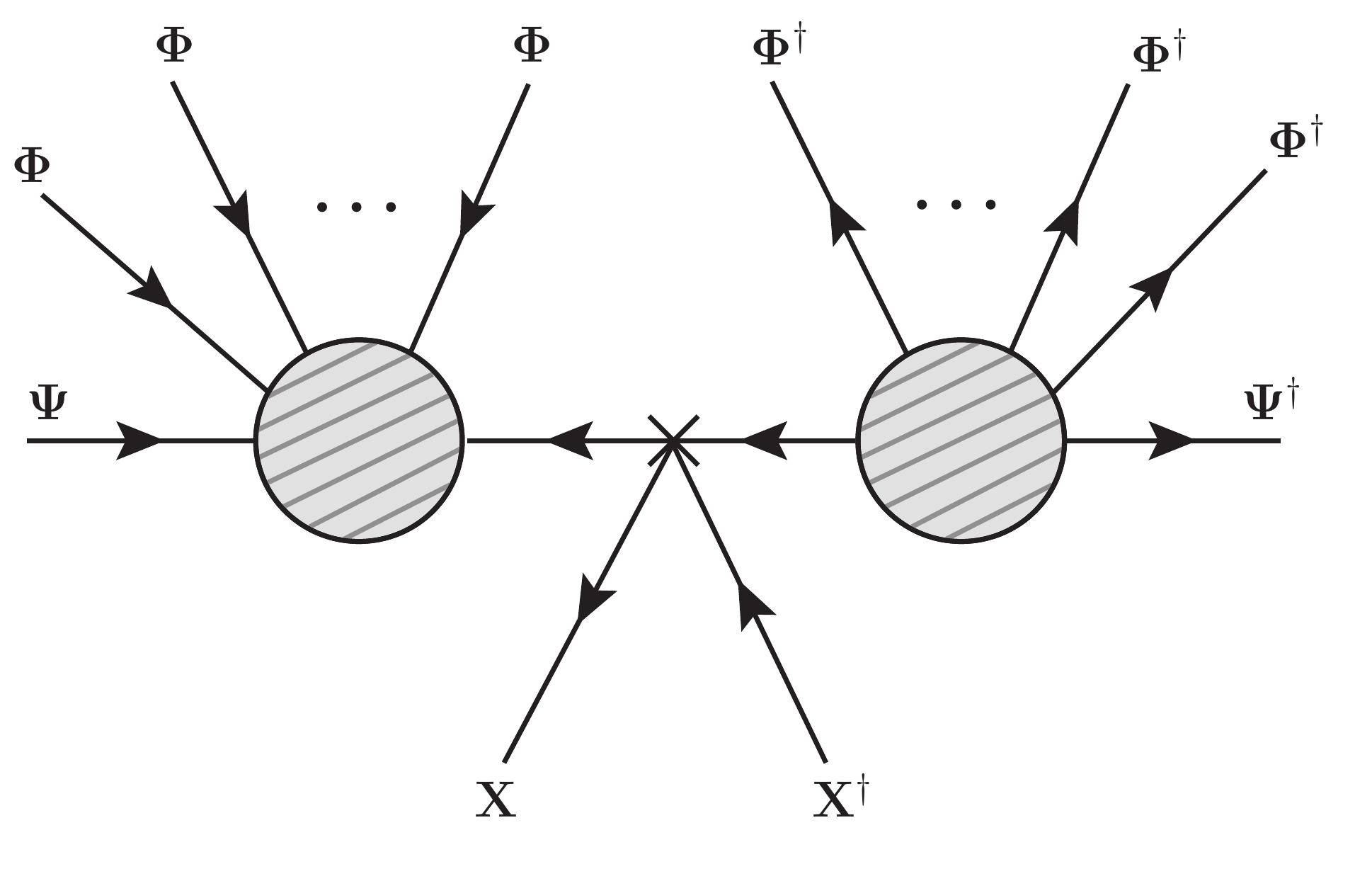}}
     \end{minipage}
 \subfigure[]{\includegraphics[width=1\textwidth]{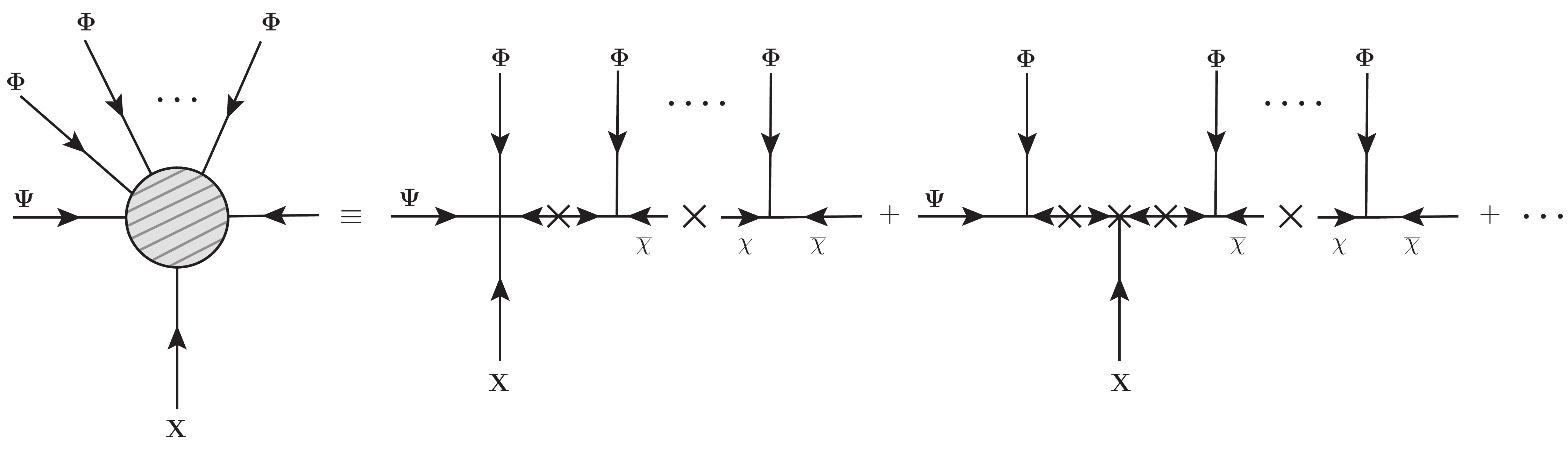}}
\caption{{\rm (a)} and {\rm (b)}: supergraph representations of the corrections to the soft mass couplings, second term in Eq.~\ref{eqn:Lsoft}. Each of the spurion fields $X$ and $ X^\dagger$, can be attached to any of the vertices or internal lines inside a blob as shown in {\rm (c)} in $(2\,n-1)$ ways, with $n$ the number of entering or outgoing fields. Moreover the two spurions can also be attached to the internal line connecting two blobs {\rm (b)}. This results in $[(2\, n_{\rm in} - 1)\, (2\, n_{\rm out} - 1) \;+\; 1]$ diagrams.} \label{fig:softmass}
\end{figure}

Similar considerations hold for the K\"{a}hler potential and the soft-breaking masses. Figure \ref{fig:kahler} represents a leading correction to the K\"{a}hler potential coming from the non-renormalizable operators in \eq{eqn:kahler}. This contribution can be schematically pictured as two bubbles (groups) of $n_{\rm in}$ fields entering and $n_{\rm out}$ daggered-fields leaving connected by a heavy messenger.
As before, a given supergraph of this form will generate the soft masses for the scalar components when coupled to the spurion combination $\langle F_{X}\rangle\langle F^\dagger_{X}\rangle$. Two possibilities contribute at the same order: either ${X}$ attached to one of the incoming vertices and ${X}^\dagger$ to one of the outgoing vertices, Figure \ref{fig:softmass} (a), or both in the internal propagator, Figure \ref{fig:softmass} (b). As can be checked in Figure \ref{fig:softmass} (c), for each bubble there will be $(2\, n_{\rm in/out} -1)$ possibilities, so that, for a single universal $F_X$, the elements $({\cal K}_\Psi)_{ij}$ and $(m^2_{\Psi})_{ij}$ will be related as:

\beq \label{eqn:factors2}
(m^2_{\Psi})_{ij} ~ \propto ~[(2\, n_{\rm in} - 1)\, (2\, n_{\rm out} - 1) \;+\; 1]\; m_0^2\; ({\cal K}_\Psi)_{ij}.
\eeq
Eqs. (\ref{eqn:factors1}) and (\ref{eqn:factors2}) are very useful, since they allow to calculate the missmatch factors without knowing the exact underlying theory from the number of flavon insertions or, equivalently, the order of the operator behind each Yukawa. Once this is done, rotations of the fields should be performed to study the phenomenology, first to canonically normalize the K\"{a}hler metric and then to the fermion mass basis.

\section{A $\boldsymbol{\Delta(27)}$ model for quarks and leptons}
\label{sec:model}

$\Delta(27)$, a finite subgroup of $SU(3)$, has been extensively studied as a flavor symmetry in GUT models, due to it being one of the smallest finite groups with triplet and anti-triplet irreducible representations.
In addition to the first work using the group as a flavor symmetry, \cite{deMedeirosVarzielas:2006fc} (considered in \cite{Lopez-Ibanez:2017xxw}), and the model we consider here \cite{deMedeirosVarzielas:2017sdv}, $\Delta(27)$ has been used in unified models \cite{Bazzocchi:2009qg, Howl:2009ds, Varzielas:2012ss, Bjorkeroth:2015uou, Bjorkeroth:2016lzs, CarcamoHernandez:2017owh}.

The models in \cite{deMedeirosVarzielas:2006fc, Bazzocchi:2009qg} are excluded by the measured value of $\theta_{13}$. The main differences between \cite{deMedeirosVarzielas:2006fc, Bazzocchi:2009qg, Varzielas:2012ss, Bjorkeroth:2015uou} and \cite{deMedeirosVarzielas:2017sdv} are most visible in the neutrino sector of the respective models, which do not significantly affect the FV constraints we consider here. Even though the models generically share similar charged fermion mass structures arising from specific VEV directions (in particular, the $(1,1,1)$ direction as a flavor symmetry breaking VEV), the subtle differences in the vacuum alignment of the respective models are relevant, as they can significantly alter the FV constraints.

We now review the relevant details of the model in Ref.\cite{deMedeirosVarzielas:2017sdv}, where the interested reader can find a more complete description.
Although the model is compatible with an underlying $SO(10)$ grand unification, we present the matter superfields as separate Left (LH), $\psi$, and Right-Handed conjugate (RH), $\psi^c$, fermions. The flavon fields, $\theta_i$, are singlets under the SM group and charged under the flavor symmetry. The model includes a Georgi-Jarlskog field $\Sigma$, associated to the breaking of the GUT symmetry, distinguishing down-quark and charged-lepton Yukawas, and a flavor singlet $S$, needed to preserve the texture zero in the neutrino Majorana matrix \cite{deMedeirosVarzielas:2017sdv}.
{\renewcommand{\arraystretch}{1.3}
\begin{table}[h!]
\centering 
\begin{tabular}{|c|c|c|c|c|c|c|c|c|c|c|}
\hline
~${\bf Field}$ & $\psi_{q,\, e,\, \nu}$ & $\psi_{q,\, e,\, \nu}^c$ & $H_5$ & $\Sigma$ &
$S$ & $\theta_3$ & $\theta_{23}$ & $\theta_{123}$ & $\theta$ & $\theta_X$\\ 
\hline
~$\Delta(27)$           & \bf 3         & \bf 3         & $\bf 1_{00}$ & $\bf 1_{00}$  & $\bf 1_{00}$ & $\bar{\bf 3}$ & $\bar{\bf 3}$ & $\bar{\bf 3}$ & $\bar{\bf 3}$  & \bf 3\\
~$Z_N$    & 0 & 0 & 0 & 2 & -1 & 0 & -1 & 2 & 0 & x \\
\hline
\end{tabular}
\caption{Transformation of the matter superfields under the $\Delta(27)$ flavor symmetryg.}
\label{tab:D27pc}
\end{table}

The field content in Table~\ref{tab:D27pc} give rise to the following superpotential, which leads to the lepton and quarks Yukawas
\bea \label{eqn:D27sp}
  {\cal W_\psi} & = & \frac{1}{M^2}\, (\psi\,\theta_3)(\psi^c\,\theta_3)\, H_5 \;+\; \frac{1}{M^3}\, (\psi\,\theta_{23})(\psi^c\,\theta_{23})\, \Sigma\, H_5 \\
  & + &  
  \frac{1}{M^3}\, (\psi\,\theta_{23})(\psi^c\,\theta_{123})\, S\, H_5 \;+\; \frac{1}
  {M^3}\,(\psi\,\theta_{123})(\psi^c\theta_{23})\, S\, H_5 
  \\
  & + & \;  
  \frac{1}{M^4}\, (\psi\,\theta_{23})(\psi^c\,\theta_{3})\, \Sigma\, S\, H_5 \;+\; \frac{1}
  {M^4}\,(\psi\,\theta_{3})(\psi^c\theta_{23})\, \Sigma \,S\, H_5 \label{eqn:D27sp3}
 \\
  & + & \;  
  \frac{1}{M^4}\, (\psi\,\theta_{3})(\psi^c\,\theta_{123})\, S^2\, H_5 \;+\; \frac{1}
  {M^4}\,(\psi\,\theta_{123})(\psi^c\theta_{3})\, S^2\, H_5\,,
  \label{eqn:D27sp4}\eea
and where the last 2 lines, Eqs.~(\ref{eqn:D27sp3}) and (\ref{eqn:D27sp4}), are comparatively suppressed and do not change the entries (12 and 21) in the matrices that are most relevant for FV bounds, so we have neglected them.

The analysis and minimization of the flavon potential carried out in Appendix A of \cite{deMedeirosVarzielas:2017sdv} aligns the VEVs of the flavons in  the directions
\bea
    \langle \theta_3 \rangle\:\propto\:\left(\begin{array}{c} 0\\ 0\\ 1\end{array}\right)  
    \hspace{4mm},\hspace{4mm}\langle \theta_{23} \rangle\:\propto\:\frac{1}{\sqrt{2}}\,\left(\begin{array}{c} 
    0\\ 1\\ 1\end{array}\right) 
    \hspace{4mm},\hspace{4mm}\langle \theta_{123} \rangle\:\propto\:\frac{1}{\sqrt{3}}\,\left(\begin{array}{c} 1\\
    1\\ 1 \end{array}\right) \,,\nn
\eea
written up to relative phases.
The VEV of the $\theta_{3}$ flavon is directly related to the third-generation of Dirac-fermions masses $y_{3}=\{y_\tau, y_t, y_b\}$ such that $\langle\theta_{3}\rangle^2/M_{a}^2 \equiv y_{3,a}$ with $a=e,u,d$. The relation between the VEVs and the parameter expansion $\varepsilon_a$ is determined by requiring a hierarchical Yukawa structure in which the 23 block is dominant with respect to the 12 block
$$  \frac{\langle\theta_{23}\rangle^2\, \langle \Sigma \rangle}{M_{23,a}^3}\, \frac{M_{3,a}^2}{\langle \theta_3\rangle^2}\,\propto\, e^{i\,\delta_a}\,r_a\,\varepsilon_a^2 \hspace{.5cm},\hspace{.5cm} \frac{\langle\theta_{23}\rangle\, \langle \theta_{123}\rangle\, \langle S \rangle}{M_{123,a}^3}\, \frac{M_{3,a}^2}{\langle \theta_3\rangle^2} \,\propto \,e^{i\,\gamma_a} \varepsilon_a^3\,,$$
where $\delta_a, \gamma_a$ are phases that appear in the mass matrices and that arise by combining the phases of the various VEVs and coefficients that contribute to the respective entries. The resulting mass matrices are complex and the model is able to reproduce the measured CP-phase in the quark sector and additionally predicts a value for the leptonic CP-phase.

Given that $\langle \Sigma \rangle/M_a = r_a$, with $r_d=r_{u}=1/3$ and $r_e=-1$, the first condition tells us that $\langle \theta_{23} \rangle/M_a = \sqrt{y_{3,a}} \,e^{i\,\delta_a/2}\,\varepsilon_a $. On the other hand, the second relation only tells us that $\langle \theta_{123} \rangle \langle S \rangle /M_a^2 \sim \varepsilon_a^2 $ leaving some freedom in the VEV assignment and, to be general, we write $\langle \theta_{123}\rangle/M_a = \sqrt{y_{3,a}}\,e^{i\,(\gamma_a-\delta_a/2)}\,\varepsilon_a^\alpha $ and $\langle S \rangle/M_a = \varepsilon_a^{2-\alpha}$ with $\alpha \in [0,1]$.

Given the above considerations, at LO the resulting Yukawa is written as 

\beq \label{eqn:D27YA}
 Y_a = y_{3,a}\: \left(\begin{array}{ccc}
              0 \hspace{.5cm}& x_{1,a}\,e^{i\,\gamma_a}\,\varepsilon_a^3 \hspace{.5cm}& x_{1,a}\,e^{i\,\gamma_a}\,\varepsilon_a^3 \\[5pt]
              x_{1,a} \,e^{i\,\gamma_a} \,\varepsilon_a^3 \hspace{.5cm}& x_{2,a}\,r_a\,e^{i\delta_a}\,\varepsilon_a^2 \hspace{.5cm}& x_{2,a}\,r_a\,e^{i\delta_a}\,\varepsilon_a^2 \\[5pt]
              x_{1,a}\,e^{i\,\gamma_a}\,\varepsilon_a^3 \hspace{.5cm}& x_{2,a}\,r_a\,e^{i\delta_a}\,
              \varepsilon_a^2 \hspace{.5cm}& 1   
             \end{array} \right) \,,
             \\
\eeq
independent of the value of $\alpha$, which on the other hand is important in the soft mass terms. In fact, the present analysis has been carried out considering three reference values for $\alpha=\{0,1/2,1\}$. However, the minimization of the flavon potential, which contemplates the case in which the self-coupling terms for the $\theta_{3,123}$ fields are dominant, facilitates large values of $\theta_{123}$, and we stress that $\alpha\in [0,1/2]$ is more consistent with this hypothesis.

It is clear that no operator in Eq.(\ref{eqn:D27sp}) can contribute to the (1,1) element, which gives the (1,1) texture zero in all mass matrices that characterizes this model (in contrast, several earlier models had the (1,1) texture zero but only in the mass matrices of the charged fermions \cite{deMedeirosVarzielas:2005ax}, \cite{deMedeirosVarzielas:2006fc, Varzielas:2012ss}). As all fermions and additional flavons are triplets under the finite group, the Dirac masses of both the quarks and leptons share the same universal form, given by Eq.(\ref{eqn:D27YA}).

The different hierarchy in the up and down sectors requires two different expansion parameters, namely $\varepsilon_u\sim\varepsilon_d/3$. This cannot be achieved if the  messengers $\chi$ are $SU(2)_L$ doublets coupling equally to $(u,d)_L$ so these messengers should be considered much heavier than their singlet counterparts. The singlet messengers in the up-sector can then be taken slightly heavier to accommodate the required difference. Beyond these differences, we consider each type of messenger to have universal masses, denoted $M$ or $M_a$ below, although we note that the messenger masses for each term can in general be different (denoted as $M_{3,a}$, $M_{23,a}$, $M_{123,a}$).

{\renewcommand{\arraystretch}{1.3}
\begin{table}[h!]
\centering 
\begin{tabular}{|c|c|c|c|c|c|c|}
\hline
~ {\bf Param.}& $\varepsilon_{e,d}$ & $(x_1,x_2)_{e,d}$ & $(\gamma,\delta)_{e,d}$ & $\varepsilon_u$ & $(x_1,x_2)_u$ & $(\gamma,\delta)_u$ \\
\hline
~{L.O.}& $0.15$  & $(1.24\,,\,2.42)$ & $(0.13\,,\,1.83)$ & $0.05$ & $(-1.12\,,\,3.60)$  & $(0\,,\,0)$ \\
\hline
~{H.O.}& $0.15$  & $(1.23\,,\,2.52)$ & $(0\,,\,2)$ & $0.05$ & $(-1.12\,,\,3.30)$  & $(0\,,\,0)$ \\
\hline
\end{tabular}
\caption{Order-one parameters from the fit of fermionic masses and mixings in \cite{deMedeirosVarzielas:2017sdv}.}
\label{tab:coeff}
\end{table}

In Ref.~\cite{deMedeirosVarzielas:2017sdv} a detailed numerical fit was performed considering both only leading order (L.O.) terms or including higher order (H.O.) corrections, and it was founf that the present measurements of the fermion masses and mixings in the lepton and quark sectors, can be accommodated. We summarize the results of this analysis in Table \ref{tab:coeff}.
In order to fit the model,  the experimentally allowed ranges for the observables are evolved to the high scale (see \cite{deMedeirosVarzielas:2017sdv} for details).

The L.O. fit was done without considering the terms with $\Sigma\,S$ and $S^2$ in the last two lines of $\mathcal{W}_\psi$ in Eqs.(\ref{eqn:D27sp}). It is a good fit to the observables, having a $\chi^2_{d.o.f} <1$ \cite{deMedeirosVarzielas:2017sdv}, although the values of certain CKM elements are out of the expected ranges and the resulting $\theta_{23}^q$ mixing angle is slightly too low, as the L.O. prediction $\theta_{23_{L.O.}}^q = 0.0191$ is just outside of the $3 \sigma$ range evolved up to the high scale, of $[0.0220, 0.0468]$.

The H.O. fit was done considering all the terms allowed by the symmetry, in particular the $\Sigma\,S$ terms whose contribution is larger than the contribution from the $S^2$ terms. These $\Sigma\,S$ terms contribute to the 23 and 32 entries of the Yukawa matrices, therefore distinguishing these entries from the 13 and 31 entries. This additional freedom significantly improves the fit in the quark sector, particularly as $\theta_{23_{H.O.}}^q = 0.0313$, within the $3 \sigma$ range.

\section{Analysis of FV-effects}
\label{sec:analysis}
In this section we construct the structures of the soft-breaking terms under the flavor symmetry and derive the non-proportionality factors as explained in Sec.\ref{sec:mech}. We show that performing the rotations to go to the standard basis, where the K\"{a}hler is the identity and the Yukawas are diagonal, does not diagonalize the obtained structures. 
{\renewcommand{\arraystretch}{1.5}
\begin{table}[h!]
\centering 
\begin{tabular}{|c|c|c|}
\hline
{\bf FV process} & Current Bounds    & Future Bounds     \\
\hline 
BR($\mu  \to e \gamma$)     & $4.2 \times 10^{-13}$ (MEG at PSI\cite{TheMEG:2016wtm})  & $4 \times 10^{-14}$ (MEG\,II\cite{Baldini:2013ke}) \\
BR($\mu  \to e e e$)        & $1.0 \times 10^{-12}$ (SINDRUM \cite{Bellgardt:1987du}) & ~~\quad $10^{-16}$ (Mu3e\cite{Blondel:2013ia}) \\
CR$(\mu-e\,)_{A_l}$        & - & ~~\quad $10^{-17}$ (Mu2e\cite{Bartoszek:2014mya},COMET\cite{Blondel:2013ia}) \\
BR($\tau \to e \gamma$)     & $3.3 \times 10^{-8}$ (BaBar\cite{Aubert:2009ag})  & \qquad $5\times 10^{-9}$ (Belle\,II\cite{Aushev:2010bq}) \\
BR($\tau \to \mu \gamma$)   & $4.4 \times 10^{-8}$  (BaBar\cite{Aubert:2009ag}) & \qquad $10^{-9}$ (Belle\,II\cite{Aushev:2010bq}) \\
BR($\tau \to e e e$)        & $2.7 \times 10^{-8}$  (Belle\cite{Miyazaki:2011xe}) & \qquad $5\times10^{-10}$ (Belle\,II\cite{Aushev:2010bq}) \\
BR($\tau \to \mu \mu \mu$)  & $2.1 \times 10^{-8}$ (Belle\cite{Miyazaki:2011xe})  & \qquad $5\times10^{-10}$ (Belle\,II\cite{Aushev:2010bq}) \\
\hline
$\Delta M_K$       & \multicolumn{2}{c|}{$(52.89 \pm 0.09)\times10^{8}\, \hslash\,s^{-1}$ \cite{Patrignani:2016xqp} } \\
$\epsilon_K$    & \multicolumn{2}{c|}{$(2.228 \pm  0.011)\times 10^{-3}$ \cite{Patrignani:2016xqp}} \\
\hline
\end{tabular}
\caption{\label{tab:LFVbounds}Relevant Flavor Violating (FV) processes considered in our analysis.}
\end{table}}
Although the analysis has been carried out numerically, we display analytically the resulting matrices to emphasize the order of magnitude of the off-diagonal terms responsible for large Flavor Violating effects. In particular we concentrate on LFV processes, on which we have already very restrictive bounds meant to be significantly improved in the near future as shown in Table~\ref{tab:LFVbounds} (for an updated review see Ref.\cite{Calibbi:2017uvl}). In addition, taking into account the presence of flavor-dependent phases, we must also consider flavor changing CP violating processes in the quark sector, such as $\epsilon_K$.

\subsection{Soft breaking terms}
The Yukawa and trilinears share the same overall texture but they will not be proportional \cite{Das:2016czs,Lopez-Ibanez:2017xxw} due to the different numerical factors, in this case of 7, 7 and 5 appearing due to the multiple topologies possible for the respective trilinear terms:
\beq
 A_a ~ = ~ a_0\,y_{3,a}\: \left(\begin{array}{ccc}
              0 \hspace{.5cm}& 7\,x_{1,a}\,e^{i\,\gamma_a}\,\varepsilon_a^3 \hspace{.5cm}&  7\,x_{1,a}\,e^{i\,\gamma_a}\,\varepsilon_a^3 \\[5pt]
              7\,x_{1,a} \,e^{i\,\gamma_a} \,\varepsilon_a^3 \hspace{.5cm}& 7\,x_{2,a}\,r_a\,e^{i\delta_a}\,\varepsilon_a^2 \hspace{.5cm}& 7\,x_{2,a}\,r_a\,e^{i\delta_a}\,\varepsilon_a^2 \\[5pt]
             7\, x_{1,a}\,\varepsilon_a^3\,e^{i\,\gamma_a} \hspace{.5cm}& 7\,x_{2,a}\,r_a\,e^{i\delta_a}\,
              \varepsilon_a^2 \hspace{.5cm}& 5  
             \end{array} \right) \,.\\[1pt]
 \eeq

As discussed in Sec.~\ref{sec:model} we consider the messengers that are $SU(2)_L$ doublets to be much heavier than their singlet counterparts, so that RH messengers dominate the contributions (this is in agreement with the model \cite{deMedeirosVarzielas:2017sdv}).
Thus, while the off-diagonal corrections to the LH-K\"ahler potential are negligible, the RH corrections remain relevant:
\bea
  {\cal K}_{\psi^c}  =  \psi^{c}\psi^{c\dagger} &\;+\;& \frac{1}{M^2}\left[( \psi^c \theta_3)(\theta_{3}^\dagger \psi^{c\dagger})\;+\; ( \psi^c \theta_{23})(\theta_{23}^\dagger \psi^{c\dagger}) \;+\; 
  ( \psi^c \theta_{123})(\theta_{123}^\dagger \psi^{c\dagger}) \right] \\
    & + & \frac{1}{M^3}\left[ ( \psi^c \theta_{3})(\theta_{23}^\dagger \psi^{c\dagger})\,S\, \;+\; {\rm h.c.} \right] \\
   & + &  \frac{1}{M^3} \left[ ( \psi^c \theta_{3})\,(\theta_{123}^\dagger \psi^{c\dagger})\,\Sigma \;+\; {\rm 
    h.c.}\right] \,
\\
   & + &  \frac{1}{M^4} \left[ ( \psi^c \theta_{23})\,(\theta_{123}^\dagger \psi^{c\dagger})\,\Sigma\,S^\dagger \;+\; {\rm 
    h.c.}\right] \, .\nn
\eea
The last term is relatively suppressed and has not been included in the phenomenological analysis.\\
Similar terms apply for the sfermion squared masses, but similar to what happens between the Yukawa couplings and the trilinears above, different prefactors appear due to multiple possible topologies. We have then the following structures 
\beq 
 K_{R,a} ~ = ~\mathbb{1}\,+\,y_{3,a}\: \left(\begin{array}{ccc}
              \varepsilon_{a}^{2\alpha}\hspace{.5cm}&  \varepsilon_{a}^{2\alpha} \hspace{.5cm}&   \,e^{i(\gamma_{a}-\frac{\delta_{a}}{2})}\,r_{a}\,\varepsilon_{a}^{\alpha}\,+\,\varepsilon_{a}^{2\alpha}\\[5pt]
              {\rm c.c.}  \hspace{.5cm}& \varepsilon_{a}^{2\alpha} \hspace{.5cm}& \,e^{i(\gamma_{a}-\frac{\delta_{a}}{2})}\,r_{a}\,\varepsilon_{a}^{\alpha}\,+\,\varepsilon_{a}^{2\alpha}\\[5pt]
              {\rm c.c.}\hspace{.5cm}& {\rm c.c.} \hspace{.5cm}& 1   
             \end{array} \right) \,,
\eeq
\\[5pt]
\beq 
 m_{R,a}^{2} ~ = ~m_0^2\,\mathbb{1}\,+\,m_0^2\,y_{3,a}\: \left(\begin{array}{ccc}
              2\,\varepsilon_{a}^{2\alpha}\hspace{.5cm}&  2\,\varepsilon_{a}^{2\alpha} \hspace{.5cm}&   4\,e^{i(\gamma_{a}-\frac{\delta_{a}}{2})}\,r_{a}\,\varepsilon_{a}^{\alpha}\,+2\,\varepsilon_{a}^{2\alpha}\\[5pt]
             {\rm c.c.} \hspace{.5cm}& 2\,\varepsilon_{a}^{2\alpha} \hspace{.5cm}&\,4\,e^{i(\gamma_{a}-\frac{\delta_{a}}{2})}\,r_{a}\,\varepsilon_{a}^{\alpha}\,+2\,\varepsilon_{a}^{2\alpha}\\[5pt]
              {\rm c.c.}\hspace{.5cm}& {\rm c.c.} \hspace{.5cm}& 1   
             \end{array} \right) \,.
\eeq
Note that even if the 12 block appears to be simultaneously diagonalized here, the rescaling of the K\"{a}hler and the rotation to the mass basis, as detailed in Appendix~\ref{app:canonical}, will re-introduce the off-diagonal terms in the soft-mass matrices.

After performing the transformations to the diagonal-Yukawa canonical basis, we obtain the following approximate form of the CKM
\beq
V_{\rm CKM}= ~ \left(\begin{array}{ccc}
              1 -\cfrac{x_{1,d}^{2}}{2\,r^{2_{d}}\,x_{2_{d}}^{2}}\, \varepsilon_{d}^{2} \hspace{.5cm}&  \cfrac{x_{1,d}}{r_d\,x_{2,d}}\,\varepsilon_d \hspace{.5cm}&  x_{1,d} \,\varepsilon_d^{3} - e^{i\,(\gamma_{d}-\delta_d)} \cfrac{r_{u}\,x_{1,d}\,x_{2,u}}{r_{d}	,x_{2,d}}\,\varepsilon_d^{2}\,\varepsilon_u \\[5pt]
              - \cfrac{x_{1,d}}{r_d\,x_{2,d}}\,\varepsilon_d\hspace{.5cm}&   1-\cfrac{x_{1,d}^{2}}{2\,r^{2_{d}}\,x_{2_{d}}^{2}}\, \varepsilon_{d}^{2}\hspace{.5cm}&   r_d\,x_{2,d}\,\varepsilon_d^{2}\\[5pt]
     	     - e^{i\,\delta_d} \cfrac{r_{u}\,x_{1,d}\,x_{2,u}}{r_{d}	,x_{2,d}}\,\varepsilon_d\,\varepsilon_u^{2}\hspace{.5cm}&  - r_d\,x_{2,d}\,\varepsilon_d^{2} \hspace{.5cm}&  1 \\[5pt]  
             \end{array} \right) \, .
\eeq
In can be checked that this matrix reproduces  to a good approximation the numerical results obtained in Ref.\cite{deMedeirosVarzielas:2017sdv}. The diagonalized Yukawas are
\beq 
\label{eq:Ydiag}
 Y_{a}^{({\rm diag})} ~ = ~ y_{3,a}\: \left(\begin{array}{ccc}
              \pm\cfrac{x_{1,a}^{2}}{r_{a}\,x_{2,a}}\,\varepsilon_{a}^{4} \hspace{.5cm}& 0 \hspace{.5cm}& 0 \\[5pt]
              0 \hspace{.5cm}& \,r_{a}\,x_{2,a}\,\varepsilon_{a}^{2}\hspace{.5cm}& 0\\[5pt]
              0 \hspace{.5cm}& 0 \hspace{.5cm}& 1   
             \end{array} \right) \,.
\eeq
To obtain the trilinear and sfermion mass matrices in this basis, we perform the following rotations:
\beq
A_{a}\rightarrow P_{L,a}^{*}\,V_{L,a}^{\dagger} \, A_{a} \, U_{R,a}\,P_{R,a}\, \hspace{1cm},\hspace{1cm} m_{R,a}^{2}\rightarrow P_{R,a}^{*}\,U_{R,a}^{\dagger} \, m_{R,a}^{2} \, U_{R,a}\,P_{R,a}\,,
\eeq
with the rephasing and rotation matrices obtained in Appendix \ref{app:canonical}. Then,
in the charged lepton sector at LO we obtain
\beq 
\label{eq:Ae}
 A_{e} ~ \rightarrow ~ a_0\,y_{\tau}\: \left(\begin{array}{ccc}
              -7\,\cfrac{x_{1,e}^{2}}{r_{e}\,x_{2,e}}\,\varepsilon_{e}^{4} \hspace{.5cm}\hspace{.5cm}& 0 \hspace{.5cm}& 0 \\[5pt]
              0 \hspace{.5cm}& -7\,\,r_{e}\,x_{2,e}\,\varepsilon_{e}^{2}\hspace{.5cm}& 2\,e^{i\,\delta_{e}}\,r_{e}\,x_{2,e}\,\varepsilon_{e}^{2}\\[5pt]
              0 \hspace{.5cm}& -2\,\,r_{e}\,x_{2,e}\,\varepsilon_{e}^{2} \hspace{.5cm}& 5   
             \end{array} \right) \,,
\eeq
\\[0.1cm]
\beq 
\label{eq:m2Re}
 m_{R,e}^{2} ~ \rightarrow ~m_0^2\,\mathbb{1}\,+\,m_0^2\,y_{\tau}\: \left(\begin{array}{ccc}
              \varepsilon_{e}^{2\alpha}\hspace{.5cm}&  -e^{2\,i\,(\gamma_{e}-\delta_{e})}\,\varepsilon_{e}^{2\alpha} \hspace{.5cm}&   3\,e^{3\,i(\gamma_{e}-\frac{\delta_{e}}{2})}\,r_{e}\,\varepsilon_{e}^{\alpha}\,+\,\varepsilon_{e}^{2\alpha}\\[5pt]
              {\rm c.c.} \hspace{.5cm}& \varepsilon_{e}^{2\alpha} \hspace{.5cm}& 3\,e^{i(\gamma_{e}-\frac{\delta_{e}}{2})}\,r_{e}\,\varepsilon_{e}^{\alpha}\,+\,\varepsilon_{e}^{2\alpha}\\[5pt]
             {\rm c.c.} \hspace{.5cm}& {\rm c.c.} \hspace{.5cm}& 1   
             \end{array} \right) \,.
\eeq
Except for the elements 12(21) and 13(31) of the trilinears, it's clear that the final matrices do not get diagonalized.  Due to the this block diagonal form of the trilinears, their relevance for FV in the lighter generations is negligible, such as in the process $\mu \rightarrow e \gamma$. The same happens in the quark sector. The absence of charge and color breaking (CCB) minima in mSUGRA\footnote{We can approximate $m_0^2\sim (m_{\tilde{\ell}}^2+ m_\ell^2 + m_{H_d}^2)$. A similar condition applies also to off-diagonal elements giving usually less stringent bounds. In the numerical analysis we put a relaxed limit on $a_0$ and discard the points not respecting the CCB condition after the RGE evolution to the EW scale is performed.}  requires $ |A_{ii}|^2\lesssim3\,m_0^2\,Y_i^{({\rm diag})}$, as shown in Ref.~\cite{Casas:1996de}. Comparing Eqs. (\ref{eq:Ae}) and (\ref{eq:Ydiag}) we see that the element $A_{22}$ gives a quite stringent bound on the allowed values of $a_0$, $a_0\lesssim \sqrt{3}\,m_0 /7$, so that the trilinear contributions are usually subdominant with respect to the soft mass matrix contributions.
Looking at Eq.(\ref{eq:m2Re}), and considering that $\alpha \in [0,1]$, we see that large off diagonal entries are obtained (we emphasize that the value of the vacuum expectation value remains perturbative, $v_{123}/M_{23,a} = \sqrt{y_3}\,\varepsilon_a^\alpha$). In particular, an $ \varepsilon_{e}^{2\alpha} \in [0.02,1] $ contribution appears in the 12 entry that controls the $\mu\rightarrow e$ LFV-decays, and $\varepsilon_{e}^{\alpha} \in [0.15,1] $ contributions arise in the 13 and 23 entries, determining the size of the LFV-transitions $\tau\rightarrow e$ and $\tau\rightarrow \mu$, respectively. Similar results are obtained in the quark sector where the resulting matrices at LO are  
\beq
 A_{u} ~ \rightarrow ~ a_0\,y_{t}\: \left(\begin{array}{ccc}
              7\,\cfrac{x_{1,u}^{2}}{r_{u}\,x_{2,u}}\,\varepsilon_{u}^{4} \hspace{.5cm}& 0 \hspace{.5cm}& 0 \\[5pt]
              0 \hspace{.5cm}& 7\,\,r_{u}\,x_{2,u}\,\varepsilon_{u}^{2}\hspace{.5cm}& 2\,e^{-i\,\delta_{d}}\,r_{u}\,x_{2,u}\,\varepsilon_{u}^{2}\\[5pt]
              0 \hspace{.5cm}& 2\,e^{i\,\delta_{d}}\,\,r_{u}\,x_{2,u}\,\varepsilon_{u}^{2} \hspace{.5cm}& 5   
             \end{array} \right) \,,
\eeq
\\[0.1cm]
\beq 
 m_{R,u}^{2} ~ \rightarrow ~m_0^2\,\mathbb{1}\,+\,m_0^2\,y_{t}\: \left(\begin{array}{ccc}
              \varepsilon_{u}^{2\alpha}\hspace{.5cm}&  -e^{i\,(\gamma_{d}-\delta_{d})}\,\varepsilon_{u}^{2\alpha} \hspace{.5cm}&   -e^{-i\gamma_{d}}\,(3\,r_{u}\,\varepsilon_{u}^{\alpha}\,+\,\varepsilon_{u}^{2\alpha})\\[5pt]
              {\rm c.c.} \hspace{.5cm}& \varepsilon_{e}^{2\alpha} \hspace{.5cm}&e^{-i\delta_{d}}\,(3\,r_{u}\,\varepsilon_{u}^{\alpha}\,+\,\varepsilon_{u}^{2\alpha}) \\[5pt]
              {\rm c.c.} \hspace{.5cm}& {\rm c.c.} \hspace{.5cm}& 1   
             \end{array} \right) \,,
\eeq
\\[0.1cm]
\beq 
 A_{d} ~ \rightarrow ~ a_0\,y_{b}\: \left(\begin{array}{ccc}
              7\,\cfrac{x_{1,d}^{2}}{r_{d}\,x_{2,d}}\,\varepsilon_{d}^{4} \hspace{.5cm}& 0 \hspace{.5cm}& 0 \\[5pt]
              0 \hspace{.5cm}& 7\,\,r_{d}\,x_{2,d}\,\varepsilon_{d}^{2}\hspace{.5cm}& 2\,r_{d}\,x_{2,d}\,\varepsilon_{d}^{2}\\[5pt]
              0 \hspace{.5cm}& 2\,e^{i\,\delta_{d}}\,\,r_{d}\,x_{2,d}\,\varepsilon_{d}^{2} \hspace{.5cm}& 5   
             \end{array} \right) \,,
\eeq
\\[0.1cm]
\beq 
 m_{R,d}^{2} ~ \rightarrow ~m_0^2\,\mathbb{1}\,+\,m_0^2\,y_{b}\: \left(\begin{array}{ccc}
              \varepsilon_{d}^{2\alpha}\hspace{.5cm}&  -e^{i\,(\gamma_{d}-\delta_{d})}\,\varepsilon_{d}^{2\alpha} \hspace{.5cm}&   3\,e^{i(2\,\gamma_{d}-\frac{3\,\delta_{d}}{2})}\,r_{d}\,\varepsilon_{d}^{\alpha}\,+\,\varepsilon_{d}^{2\alpha}\\[5pt]
              {\rm c.c.} \hspace{.5cm}& \varepsilon_{d}^{2\alpha} \hspace{.5cm}& 3\,e^{i(\gamma_{d}-\frac{\delta_{d}}{2})}\,r_{d}\,\varepsilon_{d}^{\alpha}\,+\,\varepsilon_{d}^{2\alpha}\\[5pt]
              {\rm c.c.} \hspace{.5cm}& {\rm c.c.} \hspace{.5cm}& 1   
             \end{array} \right) \,.
\eeq

\newpage
\begin{figure}[h!]
	\includegraphics[width=0.44\textwidth]{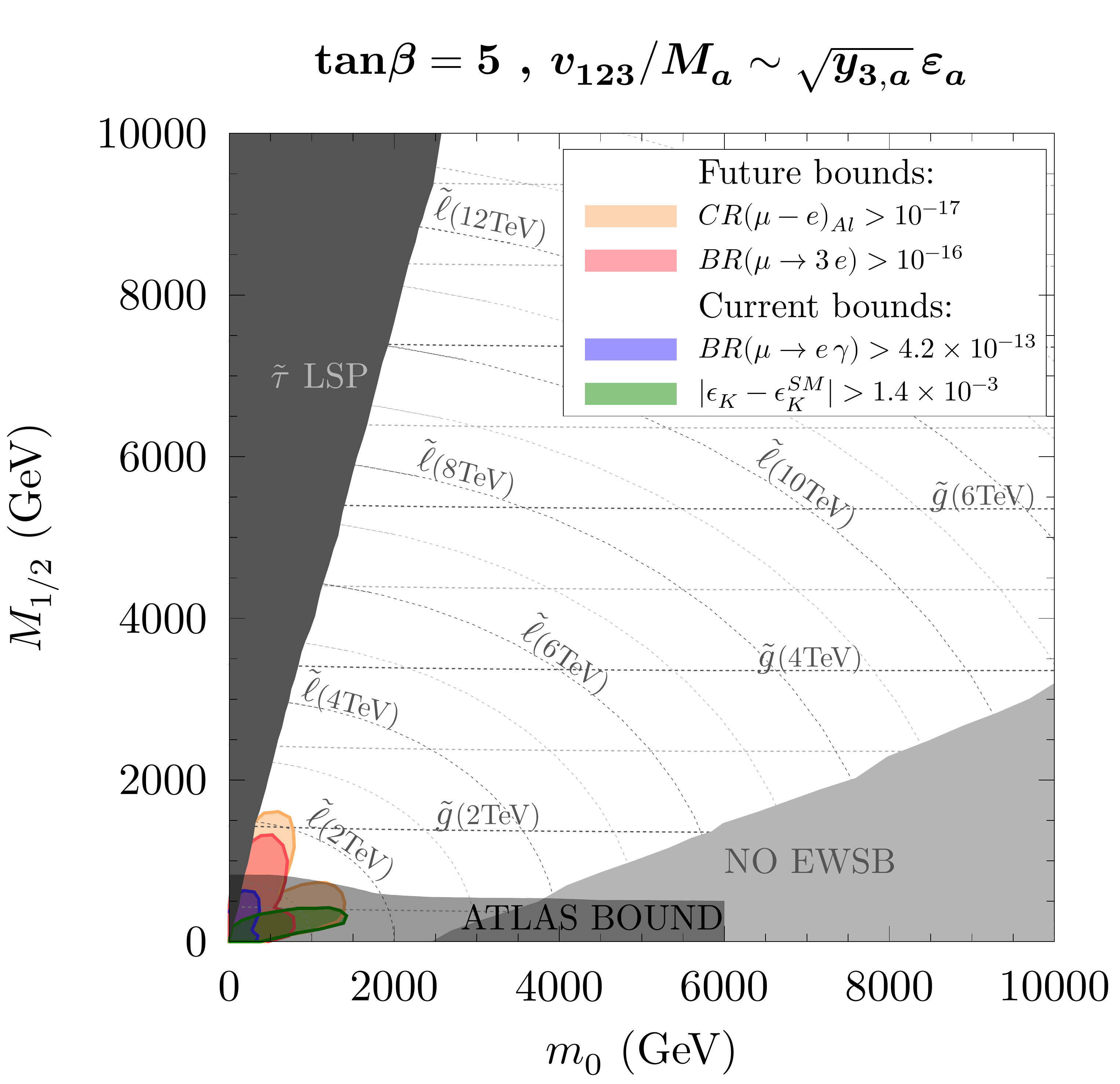}
    \includegraphics[width=0.44\textwidth]{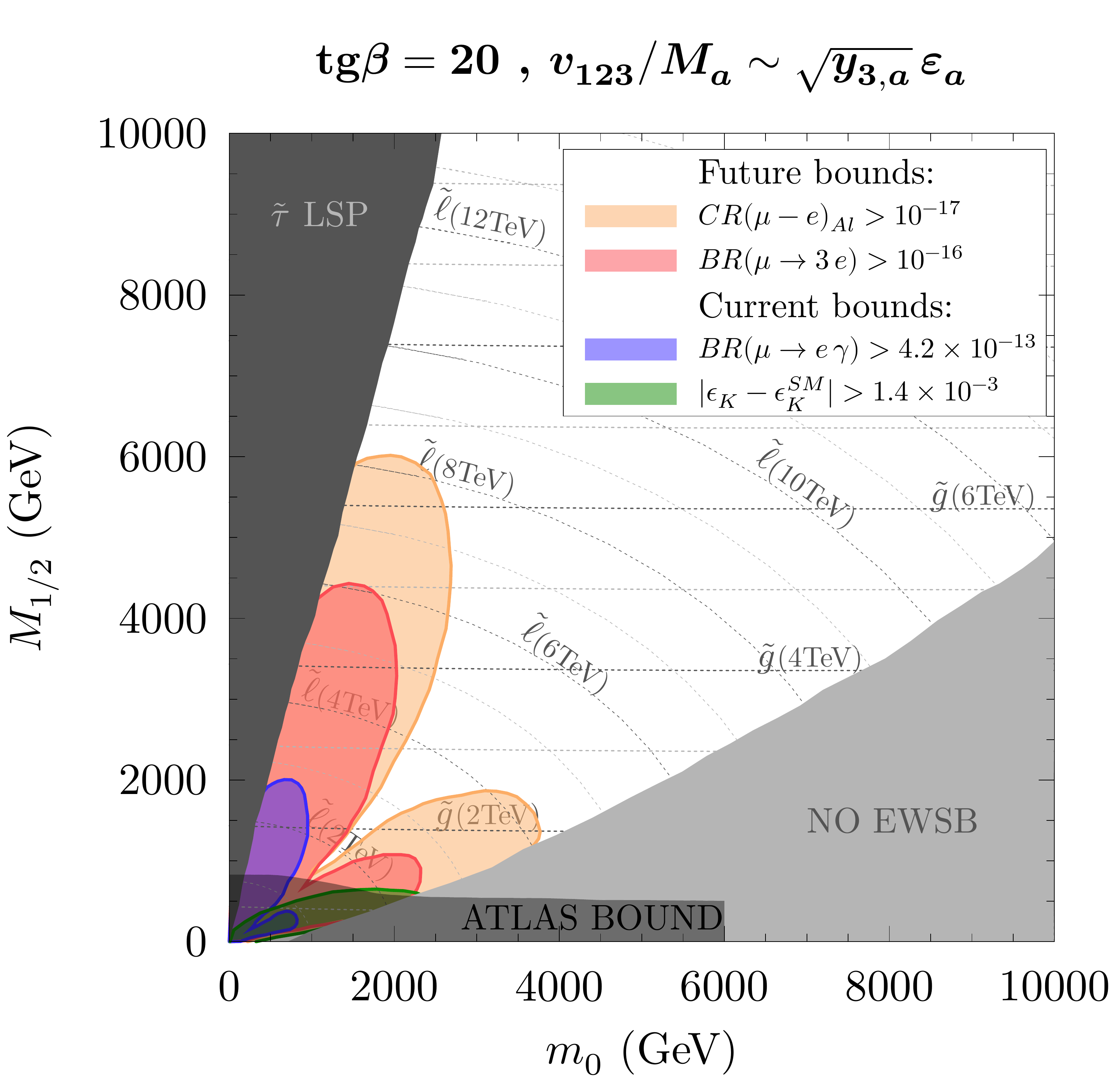}\\
    \includegraphics[width=0.44\textwidth]{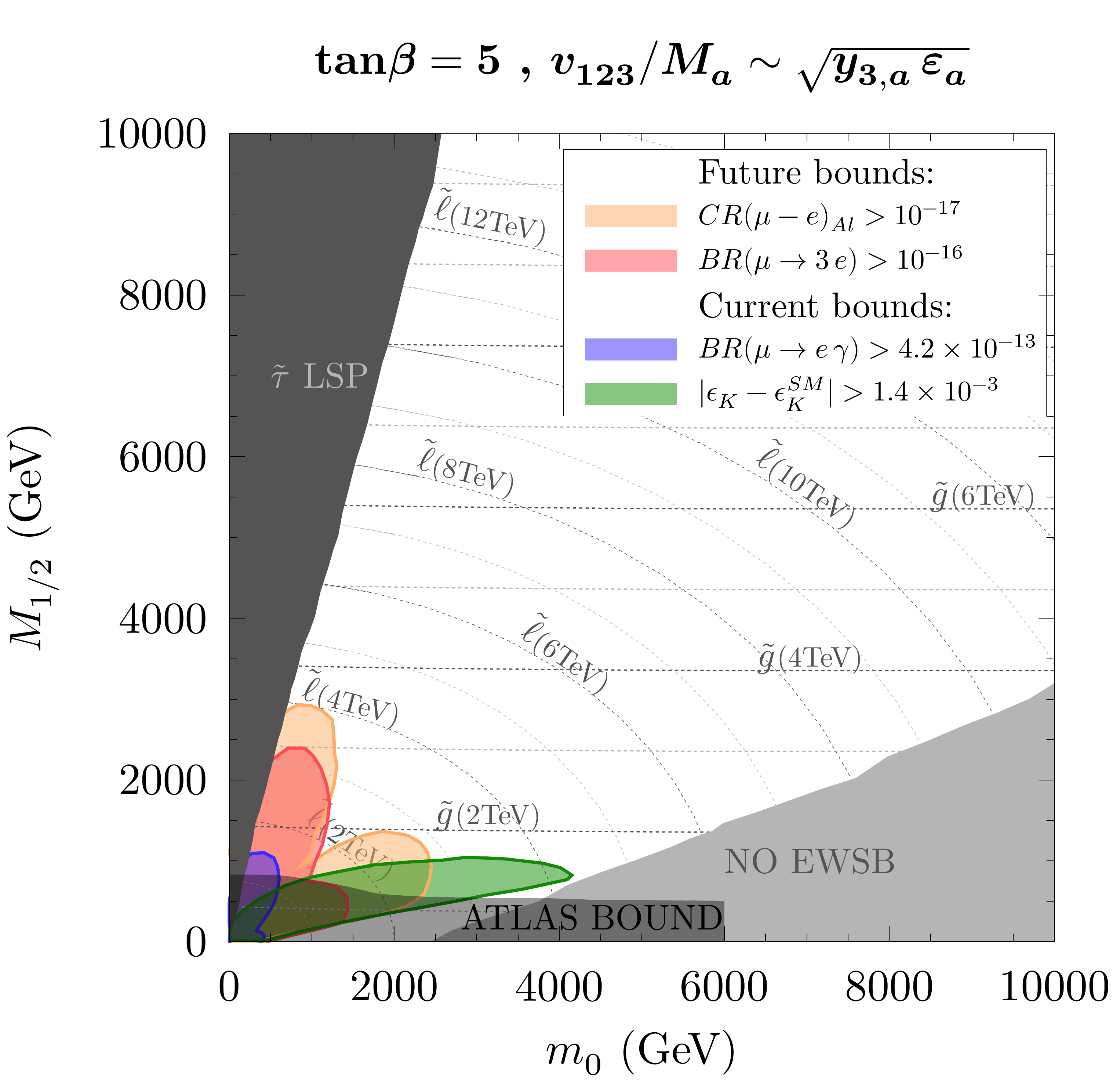}
    \includegraphics[width=0.44\textwidth]{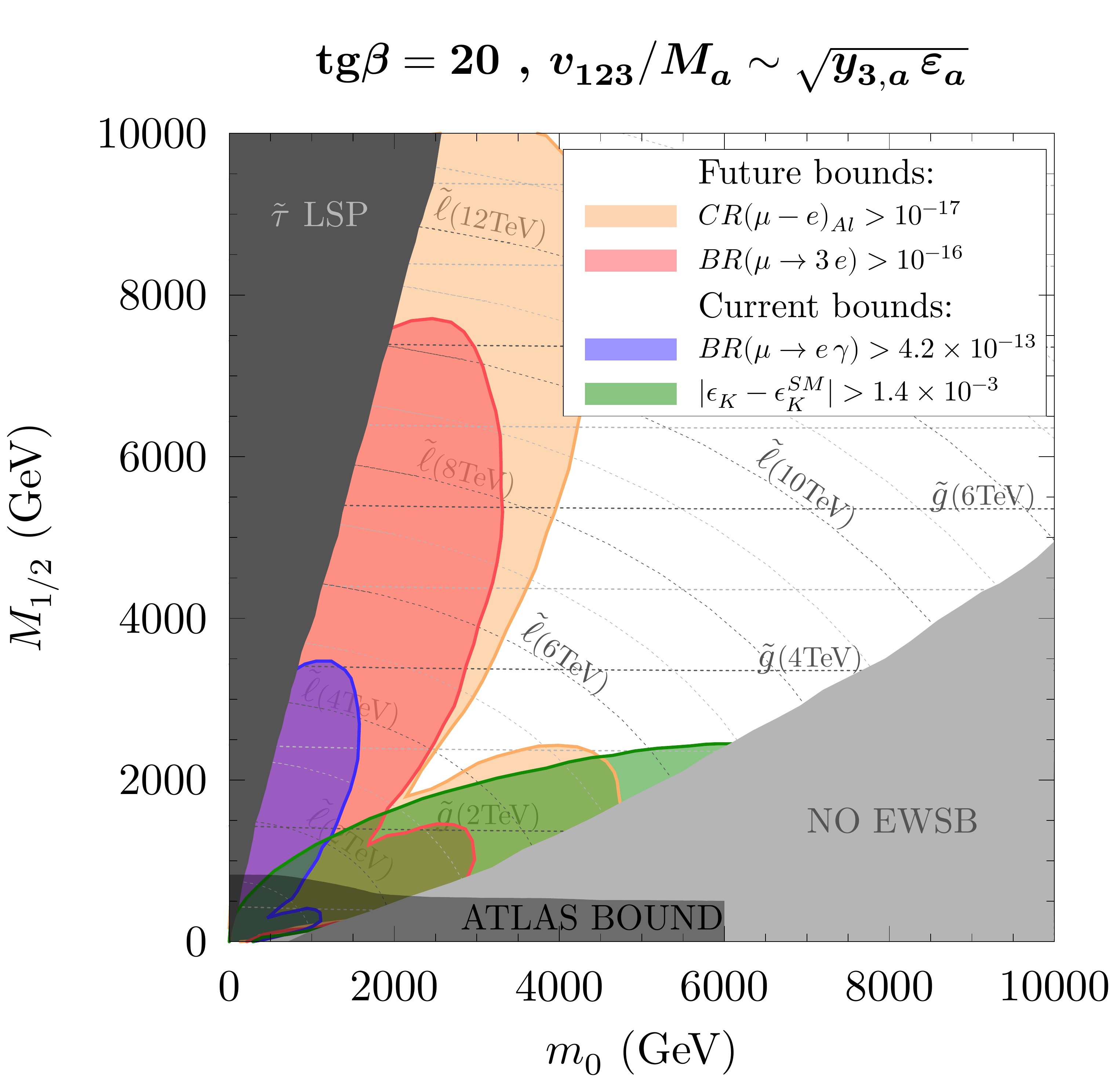}\\
    \includegraphics[width=0.44\textwidth]{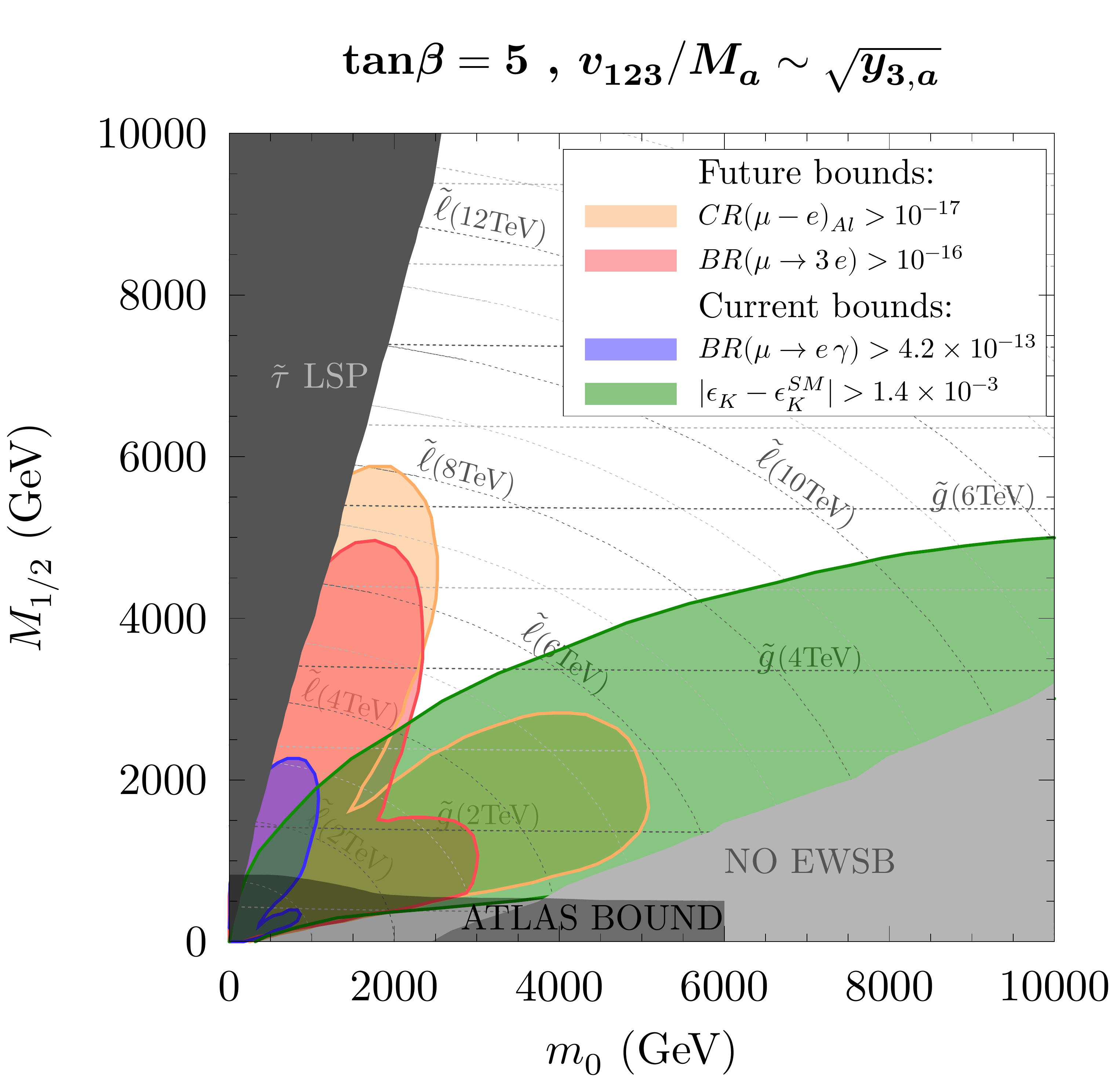}
    \includegraphics[width=0.44\textwidth]{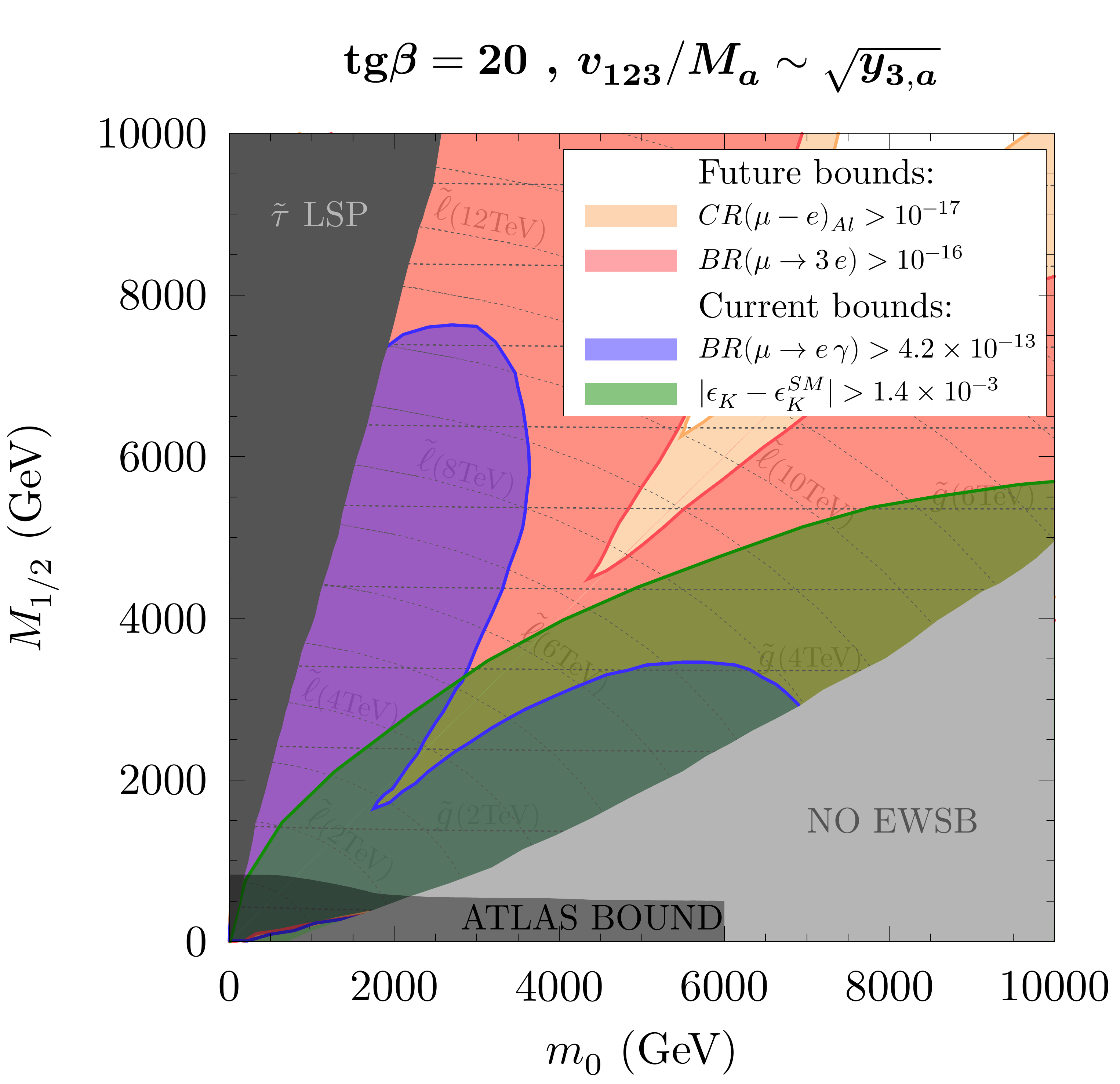}
\caption{\small Excluded regions of the MSSM parameter space due to FV constraints for the cases $\tan\beta =5,20$ and 3 different values of the VEV of the $\theta_{123}$. Blue and green shapes refer to the current bounds on $BR(\mu \rightarrow e\,\gamma)$ and $\epsilon_{K}$, red and orange shapes are the expected regions to be ruled out if future sensitivity on $BR(\mu \rightarrow 3 e)$ and $CR(\mu - e)_{Al}$ are reached with no discovery.} \label{fig:excludedregions}    
\end{figure}

\newpage 
\begin{figure}[h!]
	\includegraphics[width=0.42\textwidth]{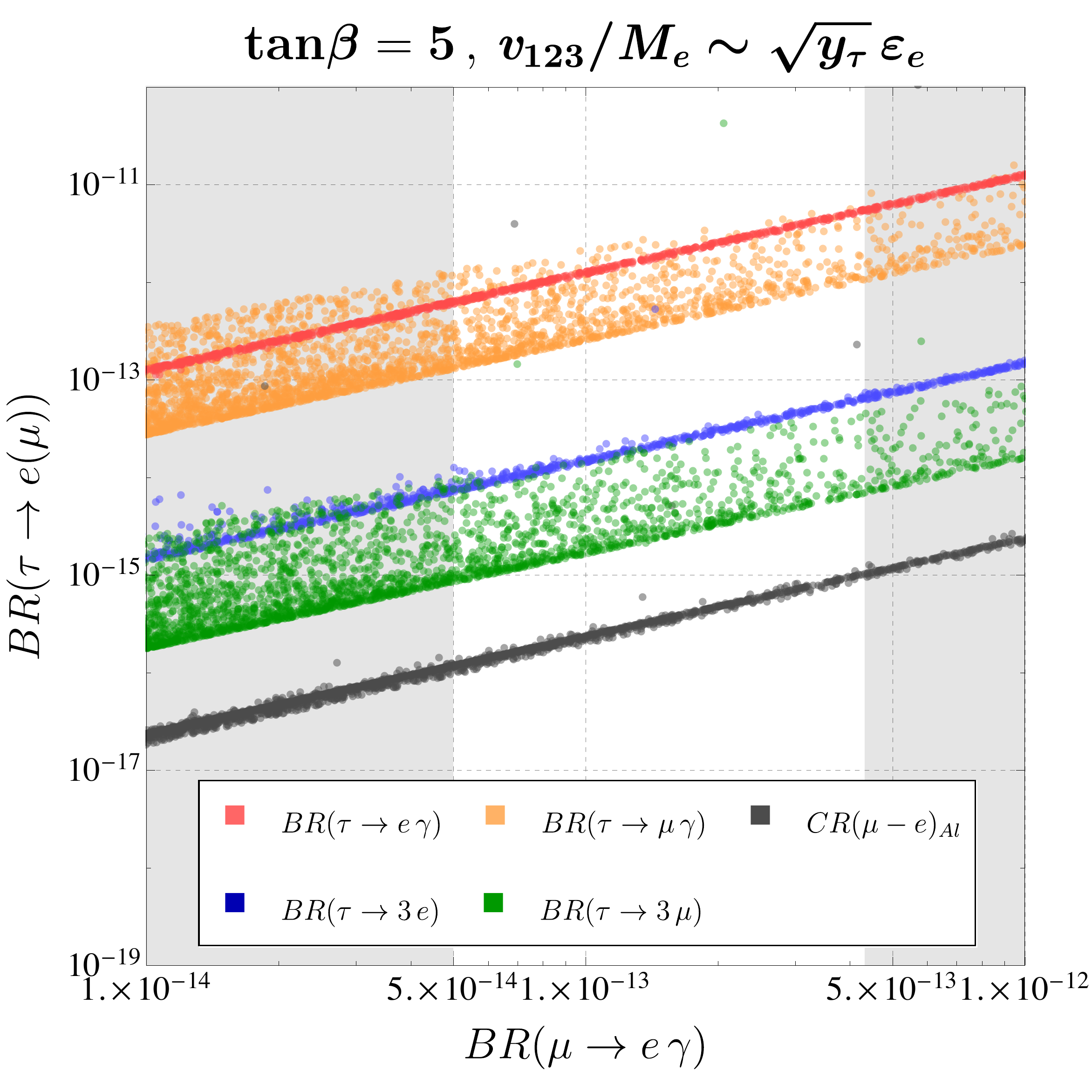}
    \hspace{1cm}\includegraphics[width=0.42\textwidth]{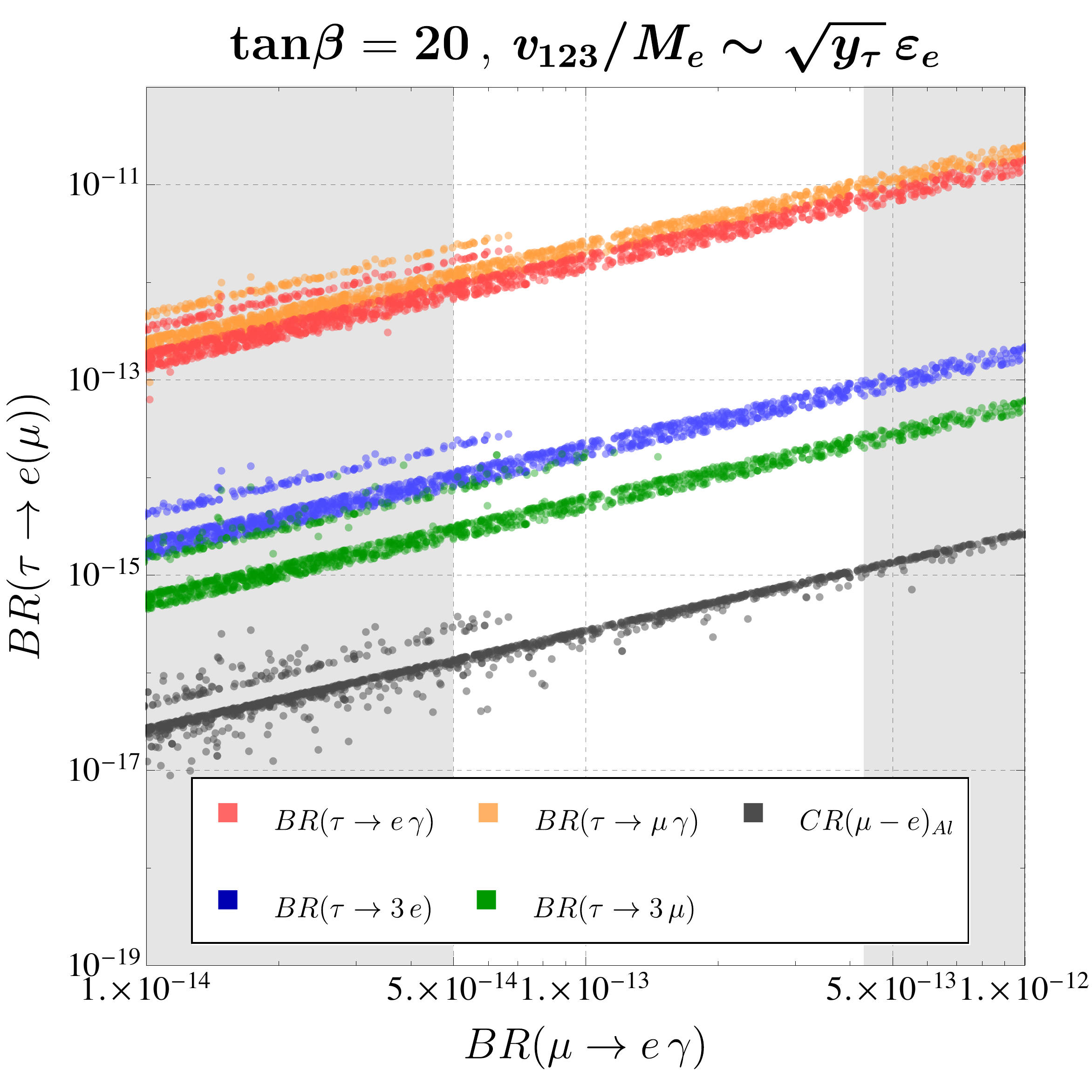}\\
    \includegraphics[width=0.42\textwidth]{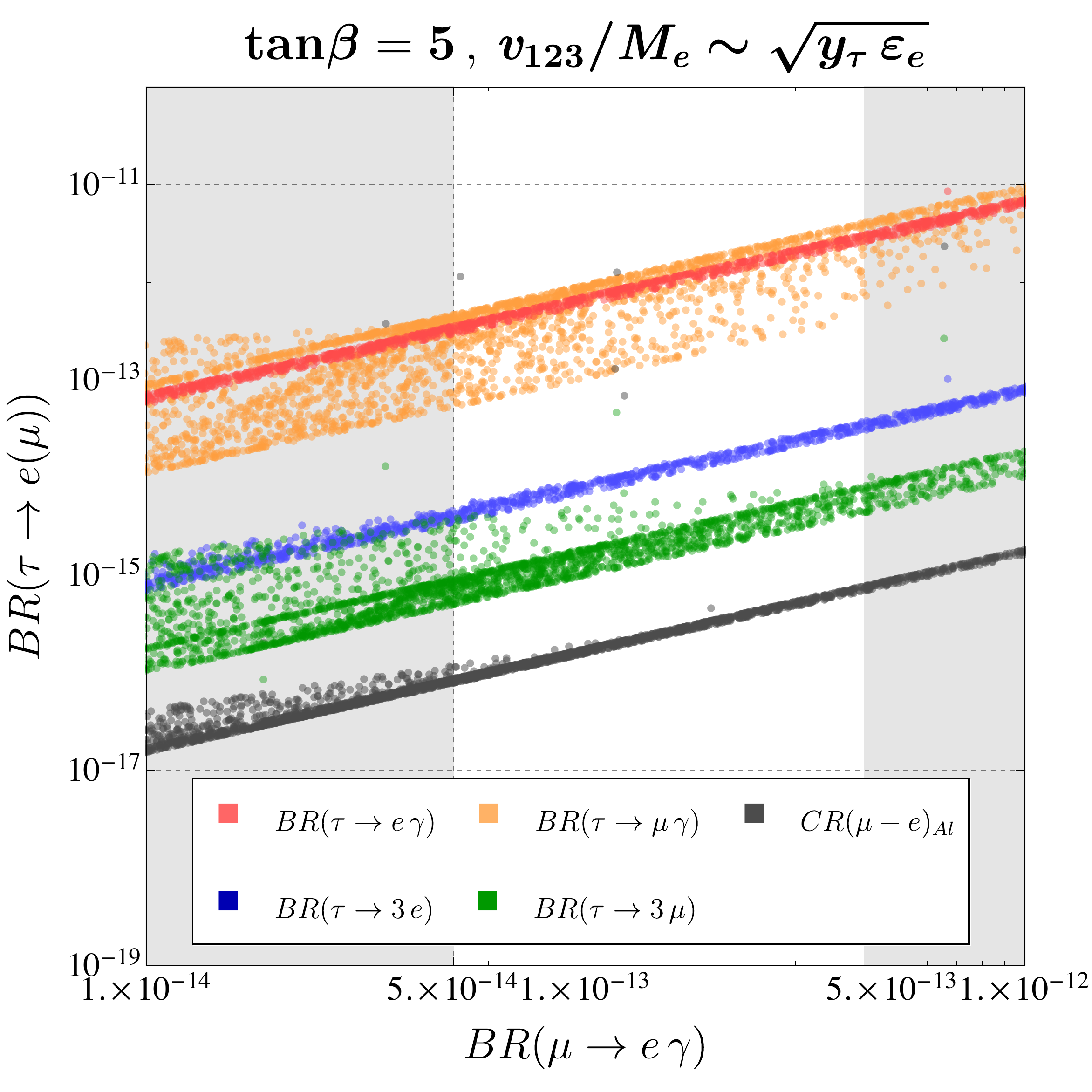}
    \hspace{1cm}\includegraphics[width=0.42\textwidth]{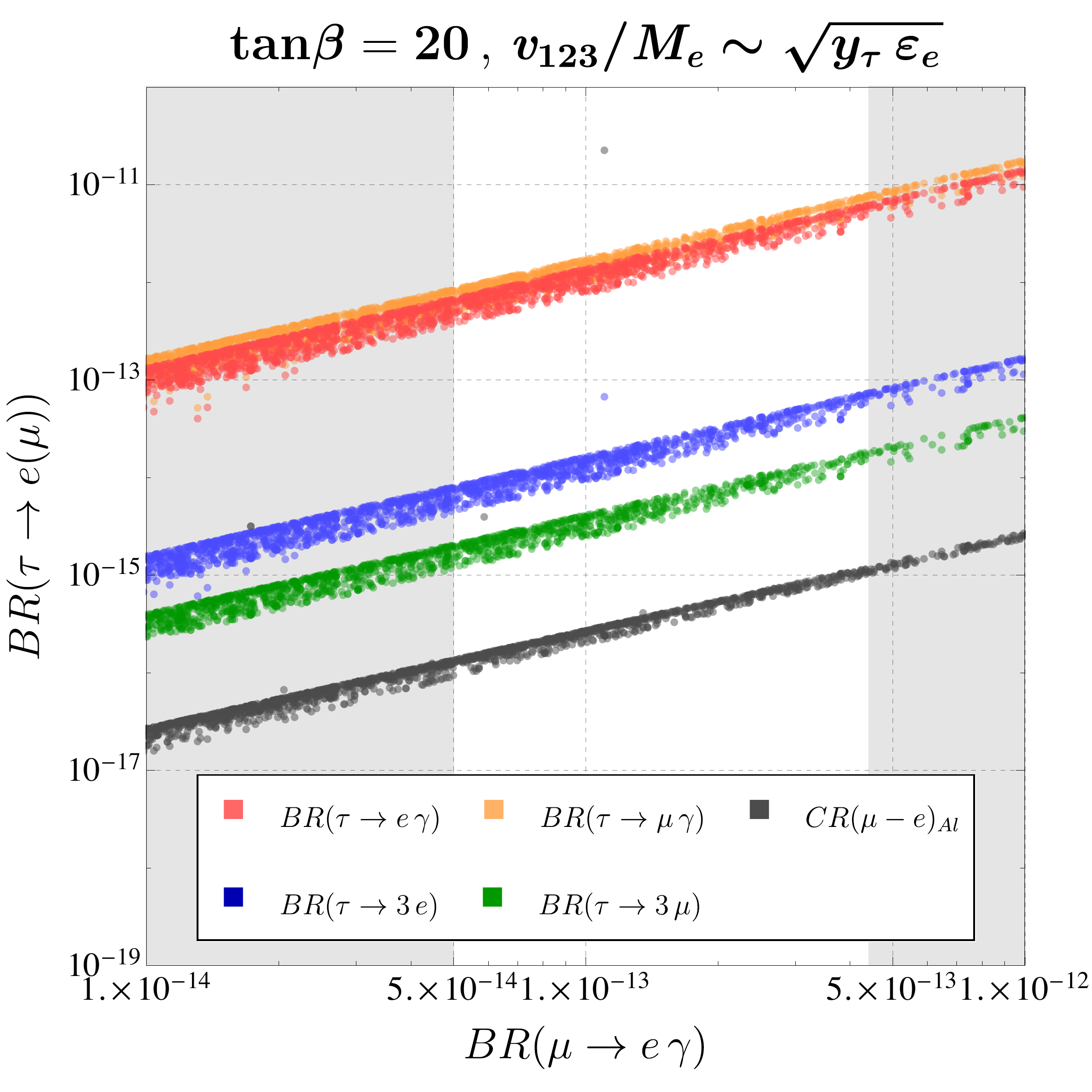}\\
    \includegraphics[width=0.42\textwidth]{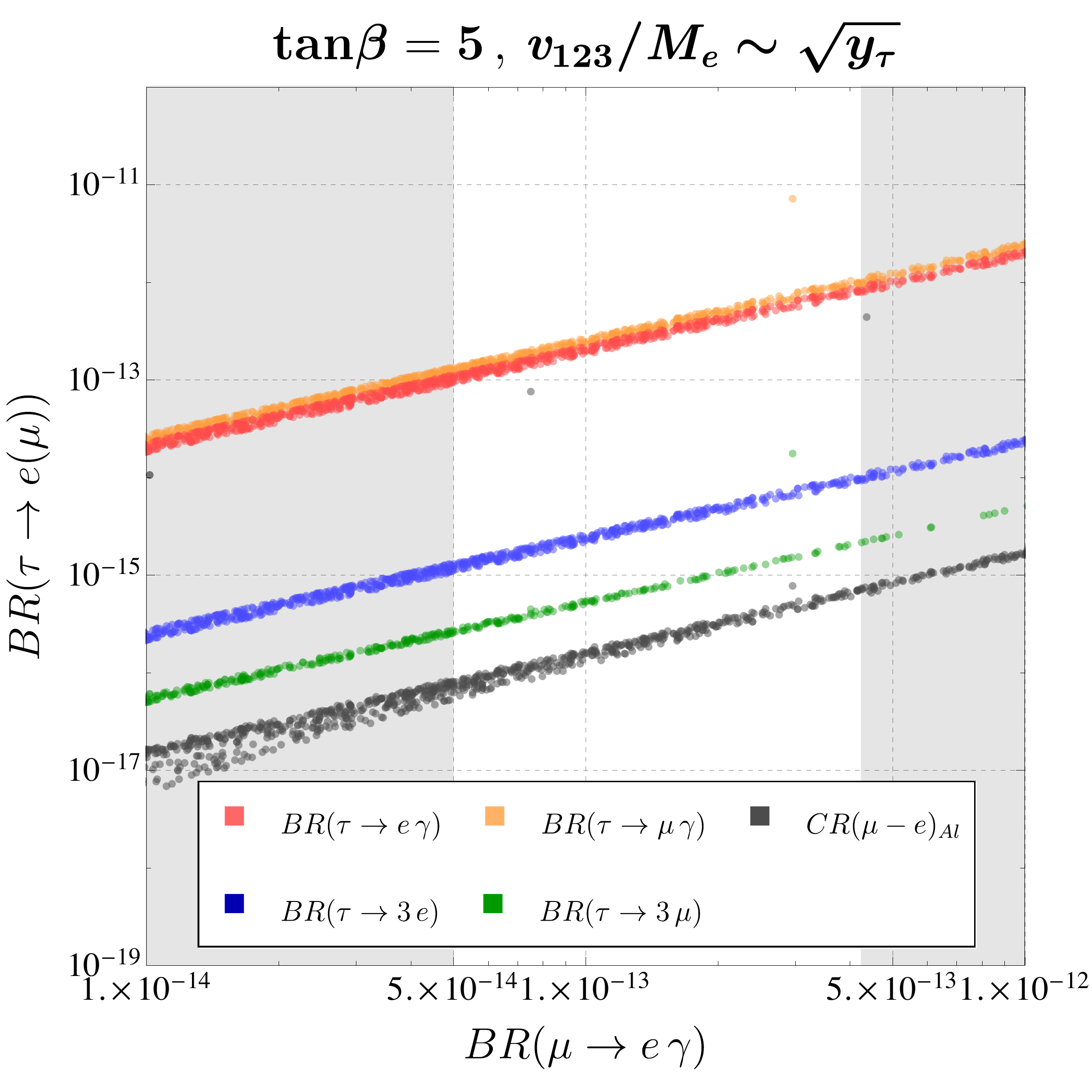}
    \hspace{1cm}\includegraphics[width=0.42\textwidth]{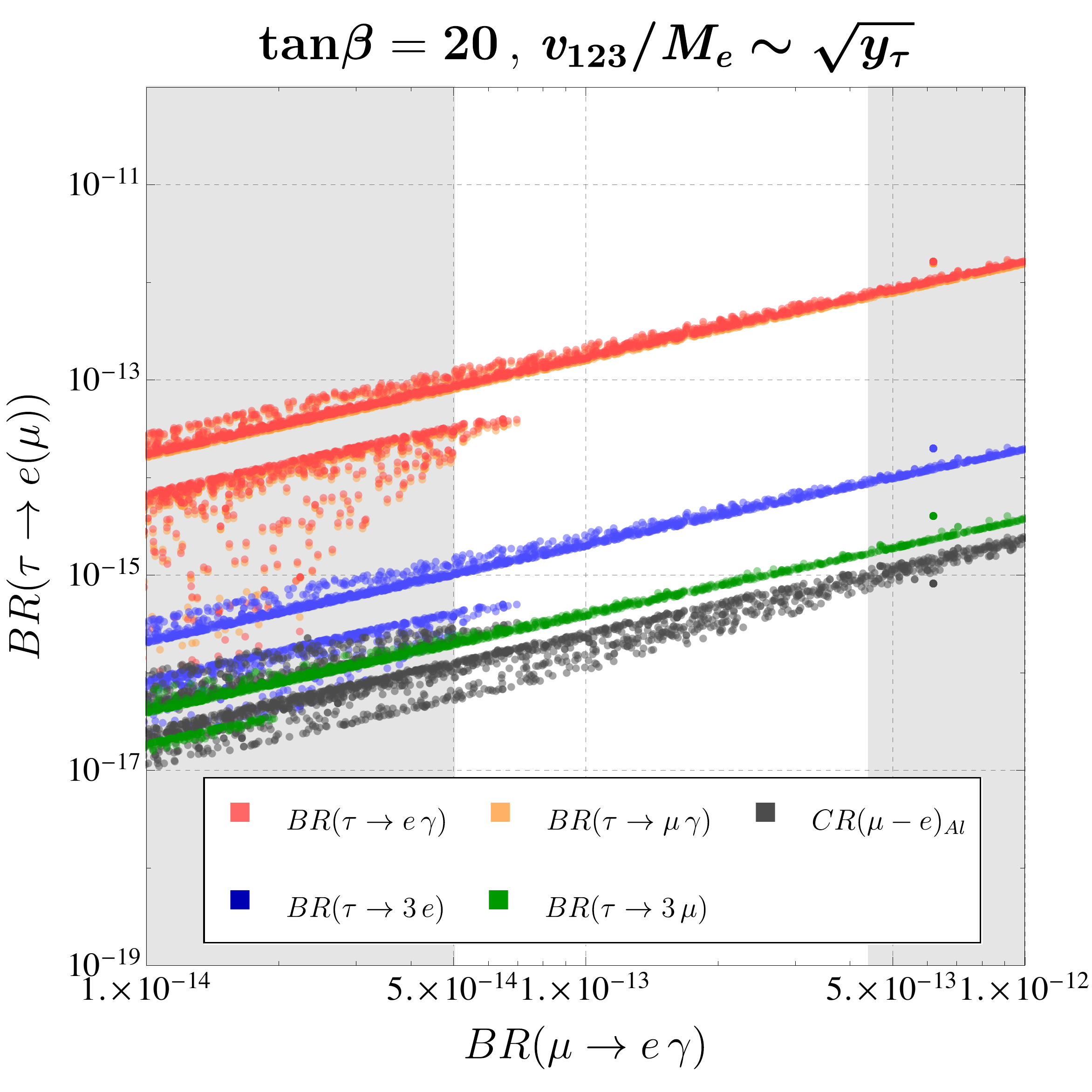}
\caption{\small Lepton flavor violating decays of the $\tau$ as a function of $BR(\mu\rightarrow\,e\,\gamma)$ for the same cases considered in Figure~\ref{fig:excludedregions}. The white window corresponds to the accessible region between the current bound and the expected future limit for $BR(\mu\rightarrow\,e\,\gamma)$, that is to the region between the blue and red shapes in Figure~\ref{fig:excludedregions}. The future limit on the $\tau$-decays is estimated as $\sim 10^{-10}$, so that the given ranges are out of reach for near future experiments.} \label{fig:taupredictions}    
\end{figure}
\newpage
In particular, we observe that the series of re-phasings that we have applied result in a dangerous $e^{\,i\,(\gamma_{d}-\delta_{d})}$ in the 12 entry of the down soft mass matrix, which can be responsible for deviations from the experimental value of $\epsilon_{K}$.
\subsection{Phenomenological results}
After substituting the numerical values corresponding to the best fit results of Table~\ref{tab:coeff}, the matrices must be evolved to the EW scale by means of the MSSM renormalization group equations (RGE), and compared to the most relevant flavor observables. Numerical calculations for the running, spectrum and low energy processes have been performed through the Supersymmetric Phenomenology package (SPheno) \cite{Porod:2003um}, together with the SARAH Mathematica package \cite{Staub:2013tta} to generate the source code. Taking into account that the flavor structures are completely fixed by the $\Delta(27)$ flavor symmetry, the only inputs are the typical supergravity parameters, chosen in the ranges: $\{m_0\,,\,M_{1/2}\} \in [0,10]$\,TeV , $a_0\in[0\,,\,0.5]\,m_0$\, and $\tan\beta=5\,,\,20$ taken as interesting representative cases.

In Fig.~\ref{fig:excludedregions}, the excluded regions of the MSSM parameter space are shown. In these plots we compare the $\tan \beta = 5$ (left) and $\tan \beta = 20$ (right) results for values of the VEV $\upsilon_{123}=\langle \theta_{123}\rangle$ (normalized with respect to the messenger mass) of $\sim \varepsilon, \sqrt{\varepsilon}$ and $1$. The blue and green shapes refer to the present bounds while the red and orange shapes are obtained from expected future limits.\\
The branching ratio of the $\ell_i\rightarrow \ell_j \gamma$ process is given by
\beq
\frac{BR(\ell_i\rightarrow \ell_j \gamma)}{BR(\ell_i\rightarrow \ell_j\,\nu_i\,\overline{\nu}_j)}=\frac{48 \,\pi^3\,\alpha}{G_F^2}(|\mathcal{A}_L^{ij}|^2+|\mathcal{A}_R^{ij}|^2)\,\sim\,\frac{\alpha^3}{G_F^2}\,\frac{\delta_{ij}^2}{m_0^4}\,\tan\beta^2\,,
\eeq
which holds approximately true for $\ell_i\rightarrow\,3\,\ell_j$ and $\mu$--$e$ conversion in atoms processes, in which the $Z$-penguin and box-type diagrams are usually subdominant with respect to the the $\gamma$-penguin $\tan\beta$-enhanced contribution.
Replacing the value of $\delta_{ij} = y_\tau \epsilon^{2\alpha}$, Eq.~(\ref{eq:m2Re}), and taking into account $y_\tau\propto \tan \beta$, we have $\tan \beta/m_0$ constant for a fixed value of the branching ratio, {\it i.e.} $\tan \beta$ scales linearly with $m_0$. Thus, in the figures, the $m_0$--$M_{1/2}$ excluded regions scale almost linearly with $\tan \beta$ (nearly a factor of 4 when going from 5 to 20). Similarly, when the VEV scales by up to a factor of $\sim 6.6$, the excluded regions growing a factor of around $4$, a milder growth in this case. 

The bounds are relatively mild in the top left corner, with small VEV and $\tan \beta = 5$, allowing $m_0$ and $M_{1/2}$ of a few TeV, and this remains true when increasing the VEV. On the other hand, even for small values of the VEV, $\tan \beta = 20$ pushes the exclusion such that typical values of $m_0$ and $M_{1/2}$ need to be larger than $4$ to $6$ TeV.
The strongest exclusions are shown in the bottom right corner, with large VEV and $\tan \beta = 20$ very little parameter space is still allowed within the displayed 10 TeV ranges for $m_0$ and $M_{1/2}$. Given such severe exclusion limits for $\tan \beta = 20$, one option is to abandon the simplifying assumption of universal messenger masses ($M_a$) and consider using the additional freedom of $M_{3,a} \neq M_{23,a}$. Together with the difference between up-type messengers and down-type messengers ($a=(u,d)$), one can accommodate simultaneously the hierarchies between $m_b/m_t$, $m_c/m_t$ and $m_s/m_b$ without going to large values of $\tan \beta$.
 
A feature of the model is the heaviness of the Left messengers which results in small off-diagonal $\delta^{LL}$-insertions. Consequently, the shapes that we observe are the typical ones of the $\delta^{RR}$ sector which is $\tan\beta$-enhanced and exhibit the usual cancellation between the bino and and bino-higgsino amplitudes in the $0.7\,m_0\lesssim M_{1/2}\lesssim 1.2\,m_0$ region of the parameter space Ref.~\cite{Paradisi:2005fk}. Such cancellations occur in all situations in which the contribution of the off-diagonal trilinear terms $\delta^{RL}$ is negligible, which is often the case considering the CCB upper bound on $|A|$ and the texture of the trilinear terms of the model where $\delta_{12(21)}\simeq\delta_{13(31)}\simeq 0$. \\ 
Apart from LFV observables, the CP-violating observable $\epsilon_K$ plays also an important role in the restriction of the  parameter space of the model. In this case, $\epsilon_K$ is, in principle, independent of $\tan \beta$, but the off-diagonal entries of squark mass matrices are proportional to $y_b$, which restores the $\tan \beta$ dependence, as we see in Fig.~\ref{fig:excludedregions}. The exclusion regions in this figure correspond to points out of the 3$\sigma$-range $|\epsilon_K - \epsilon_K^{SM}|< 3\sqrt{\sigma_{SM}^2 + \sigma_{SUSY}^2}\simeq 1.4 \times 10^{-3}$, where $\epsilon_K^{SM}$ is the SM prediction computed for each point by SPheno with only trivial soft-breaking structures as inputs and congruent with the estimate in Ref.\cite{Brod:2011ty}. The $\sigma_{SM}$, $\sigma_{SUSY}$ are the theoretical uncertainties of the SM estimate and the SUSY contribution respectively. In particular $\sigma_{SM}\sim 10\,\%\,\epsilon_K^{SM}$ \cite{Brod:2011ty} while $\sigma_{SUSY}$ is dominated by the hadronic uncertainties of the $f_K$ decay constant and the $B$-parameters coming from Lattice QCD computations (see  Ref.\cite{Ciuchini:1998ix}). To be conservative, we have taken this into account, letting  each parameter vary between its minimum and maximum value and taking half of the difference between the respective maximum and minimum values of $\epsilon_K$. We find that $\sigma_{SUSY}$ receives the largest contribution from $B_3(\mu)=1.05\,(12)$ and $B_5(\mu)=0.73\,(10)$ and can be up to $ \sim 20\,\%\,\epsilon_K^{SM}$ in the region where the SUSY contribution is comparable to the SM one. As we can see in the figure, this observable is very effective in restricting the region of low $M_{1/2}$, which corresponds to realtively light gluino and squark masses, but can reach large $m_0$ values.\\
It is interesting to compare the model predictions for LFV processes involving the $\tau$ lepton, shown in Fig.~\ref{fig:taupredictions}, with the benchmark decay $\mu \to e \gamma$. As a reflection of the $m_{R,e}^2$ structure in Eq.(\ref{eq:m2Re}), the $\tau$ branching ratios are correlated and increase linearly with the branching ratio of the benchmark decay. For the same value of $BR(\mu \to e \gamma)$, larger $\tan \beta$ (going from a plot in the left to a plot in the right) corresponds to slightly smaller $\tau$ branching ratios, while a larger VEV (going from a plot above to a plot below) corresponds to a more significant reduction of the $\tau$ branching ratios (up to one order of magnitude smaller over the the range considered).\\
It is worth noting that in the cases $\tan\beta=5$ and $\upsilon_{123}/M_e=\sqrt{y_\tau}\,\varepsilon_e,\sqrt{y_\tau\,\varepsilon_e}$, the observed dispersion for the processes $\tau\rightarrow \mu\,\gamma$ and $\tau\rightarrow 3\,\mu$ is imputable to a non-negligible effect of the $\delta_{23}^{RL}$-insertion. For larger values of $\tan\beta$ and $\upsilon_{123}$ this ceases to be the case.\\ 
In some cases, particularly in the $\tan\beta=20$ panels, for each branching ratio a second line becomes visible, and the two lines correspond to the maximum directions of growth in the $\{m_0,M_{1/2}\}$ planes of Fig.~\ref{fig:excludedregions}. This is caused by a misalignment of the cancellation region with respect to the one of $\mu\rightarrow e\,\gamma$, which results in two distinct directions of growth. The misalignment stems from additional contributions, deriving mainly from the inclusion of the two mass insertions $\delta_{ik}^{RR}\delta_{kj}^{RR}$ \footnote{In principle the misalignment could be also due to the contribution of additional diagrams, however this is not what we observe for the analyzed processes.} (see Ref.\cite{Paradisi:2005fk}), which for these processes is not negligible. Note that, as the exclusion region given by $\epsilon_K$ can reduce or exclude the points in one of the two direction of growth, we are not plotting these points in Fig.~\ref{fig:taupredictions}. 

\subsection{Comparisons with other models}

As explained in Section \ref{sec:mech}, non-universal soft-breaking terms are always expected in supersymmetric models when the flavor symmetry is broken below the SUSY-breaking mediation scale. However, different flavor symmetries give rise to different structures in the soft-breaking terms while reproducing the observed fermion masses and mixing matrices. 

This non-universality of soft-terms in supersymmetric flavor models has long been considered in the literature \cite{Nir:1993mx,Dine:1993np, Leurer:1993gy, Pomarol:1995xc, Barbieri:1995uv, Nir:1996am, Dudas:1996fe, Barbieri:1996ww, Nir:2002ah,Ross:2004qn, Antusch:2008jf, Nomura:2008gg, Calibbi:2009ja,Calibbi:2009pv,Altmannshofer:2009ne,Lalak:2010bk, Babu:2011mv, Antusch:2011sq, Calibbi:2012yj, Babu:2014sga}. In most of these works, the structure of the soft terms is fixed by the symmetry and it is simply assumed that the unknown {\cal O}(1) coefficients differ from the ones in the Yukawa matrices. In Refs.~\cite{Das:2016czs,Lopez-Ibanez:2017xxw} this non-proportionality was explicitly demonstrated and the corresponding soft-terms obtained. In \cite{Das:2016czs}, two flavor symmetries were considered, $U(1)_f$ and $SU(3)_f$, while in \cite{Lopez-Ibanez:2017xxw}, three different models were considered: $\Delta(27)$ (a different model from the one considered here), $A_4$ and $S_3$. In this section, we compare their results with the present model in order to extend the validity of our results to a wider class of models.    

In Ref.~\cite{Das:2016czs}, the kaon observables $\Delta M_K$ and $\epsilon_K$ were considered, while in \cite{Lopez-Ibanez:2017xxw} only leptonic observables were taken into account. In this work, we consider both kaon and leptonic observables and therefore we can compare the exclusion regions with the results presented in these works. In general, the regions for different models have distinct shapes, depending on the respective details of the models. For instance, the shape of the region constrained by $\epsilon_K$ in  \cite{Das:2016czs} (shown there in the $m_{\tilde g}$--$m_{3/2}$ plane) is very similar to the one we find in this work for the case of the $SU(3)$ symmetry, but distinct from the shape of the $U(1)$ model.

Regions excluded by leptonic observables in \cite{Lopez-Ibanez:2017xxw} are clearly different to the ones found in this work for the $A_4$ and $S_3$ models, due to the simultaneous presence of left-handed and right-handed mass insertions.
Interestingly, the older $\Delta(27)$ model \cite{deMedeirosVarzielas:2006fc} (now excluded due to $\theta_{13}$) produces similar exclusion shapes as the unified texture zero model \cite{deMedeirosVarzielas:2017sdv} we analyse here. This is not surprising, as the main differences between the two models arise in the neutrino sector and both models use the same flavon VEV directions, in particular the $(1,1,1)$ direction which is most relevant for FV processes as it contributes to the lighter generations (where bounds are more stringent). The particularities of the vacuum alignment mechanisms employed in the two $\Delta(27)$ models would allow them to be distinguished in FV observables. For the cases where FV is enhanced by taking $v_{123}/M \sim \sqrt{y_\tau\,\varepsilon}$ or $v_{123}/M \sim \sqrt{y_\tau}$, as preferred by the vacuum alignment mechanism, current FV bounds reach much higher exclusions, up to around $8$ TeV in the most sensitive case, as seen in the bottom right panel of Fig. \ref{fig:excludedregions}.

In summary, taking into account current leptonic and kaon FV bounds allows us to extend the excluded regions in comparison to previous works.

\section{Conclusions \label{sec:conclusion}}

We performed an analysis of quark and lepton flavor violating processes in a supersymmetric model enlarged with a $\Delta(27)$ flavor symmetry which is broken below the supersymmetry-breaking mediation scale.  
We have explicitly shown the non-universality of trilinear terms and supersymmetry soft-breaking masses, with all the SUSY breaking matrices determined in terms of $m_0$, $a_0$ and $\tan \beta$.  

FV processes allow us to explore this model up to rather heavy sparticle masses, well above the LHC reach. LFV bounds, and specially $\mu \to e \gamma$, are the most restrictive constraints in the parameter space, but thanks to the presence of flavor-dependent phases, $\epsilon_K$ plays an important role in exploring the region of low $M_{1/2}$. 

The combination of LFV proceses and $\epsilon_K$ can restrict very large values of $M_{1/2}$ and $m_0$, depending on the $\tan \beta$ value and the model vev, $v_{123}$. Indeed, if we take the typical values preferred by the vacuum alignment mechanism of the model, the constraints become particularly severe, reaching values of several TeV for $M_{1/2}$ and $m_0$. This fact, implies that assumptions such as universal messenger masses are too simple and should be abandoned, in order to relax $\tan\beta$ and still accommodate the hierarchy between the top and bottom mass.

We have compared the results for this flavor model with other  models, including a similar flavor model with a $\Delta(27)$ symmetry, and discrete symmetries such as $A_4$ and $S_3$. In general, FV processes constrain these models in different ways and lead to qualitative and even quantitative differences.

In conclusion, FV searches are able to constrain the parameter space of flavor models and even distinguish flavor models that would otherwise be hard to discriminate by solely increasing the precision of fermion masses and mixing parameters.

\appendix
\section{Canonical normalization and rotation to CKM basis}
\label{app:canonical}
To compare the MSSM contributions to the SM predictions, we have to rotate the K\"{a}hler to pass to the canonical basis, where the K\"{a}hler corresponds to the identity and the kinetic terms are canonical. This can be achieved through an upper triangular matrix $U_{T}^{\dagger}\,K_{R}\,U_{T}$ \cite{Espinosa:2004ya, King:2004tx}, with $U_{T}$ of the form
\beq
U_{T}~ = ~ \left(\begin{array}{ccc}
              1-y_{3,a}\cfrac{\varepsilon_{a}^{2\alpha}}{2} \hspace{.5cm}&  -y_{3,a}\,\varepsilon_{a}^{2\alpha} \hspace{.5cm}& -\cfrac{y_{3,a}}{\sqrt{1+y_{3,a}}}\,\left(\,e^{i(\gamma_{a}-\frac{\delta_{a}}{2})}\,r_{a}\,\varepsilon_{a}^{\alpha}\,+\,\varepsilon_{a}^{2\alpha}\right)\\[5pt]
              0 \hspace{.5cm}&  1-y_{3,a}\cfrac{\varepsilon_{a}^{2\alpha}}{2} \hspace{.5cm}& -\cfrac{y_{3,a}}{\sqrt{1+y_{3,a}}}\,\left(\,e^{i(\gamma_{a}-\frac{\delta_{a}}{2})}\,r_{a}\,\varepsilon_{a}^{\alpha}\,+\,\varepsilon_{a}^{2\alpha}\right) \\[5pt]
              0 \hspace{.5cm}& 0  \hspace{.5cm}& \cfrac{1}{\sqrt{1+y_{3,a}}}   
             \end{array} \right) \,.
\eeq 
This upper triangular canonical transformation gives only sub-leading effects on the hierarchical structures of the Yukawas and Trilinears and therefore can neglected in this qualitative discussion\footnote{The only possible exception is the rescaling of the third row and column }. On the other hand, its effect on the soft mass matrices is a general 1-unit reduction in the degeneracy coefficients. The Yukawas are diagonalized by a bi-unitary transformation
\beq 
\label{eq:Ydiag_app}
Y_{a}^{(diag)}=P_{L,a}^{*}\,V_{L,a}^{\dagger}\,Y_{a}\, U_{R,a}\,P_{R,a}\,,
\eeq
where $P_{L,a}, P_{R,a}$ are diagonal re-phasing matrices introduced to go to the SM phase conventions in the CKM matrix as shown below, while
\beq 
\label{eq:VLa}
 V_{L,a} ~ =~ U_{R,a}^* ~ =~ \left(\begin{array}{ccc}
              1 -\cfrac{x_{1,a}^{2}}{2\,r^{2_{a}}\,x_{2_{a}}^{2}}\, \varepsilon_{a}^{2}  \hspace{.5cm}& e^{i\,(\gamma_a -\delta_a)} \cfrac{x_{1,a}}{r_a\,x_{2,a}}\,\varepsilon_a \hspace{.5cm}& e^{i\,\gamma_a} x_{1,a} \,\varepsilon_a^{3} \\[5pt]
              -e^{-i\,(\gamma_a -\delta_a)} \cfrac{x_{1,a}}{r_a\,x_{2,a}}\,\varepsilon_a\hspace{.5cm}&   1 -\cfrac{x_{1,a}^{2}}{2\,r^{2_{a}}\,x_{2_{a}}^{2}}\, \varepsilon_{a}^{2} \hspace{.5cm}&  e^{i\,\delta_a} r_a\,x_{2,a}\,\varepsilon_a^{2}\\[5pt]
     	     0 \hspace{.5cm}&  -e^{-i\,\delta_a} r_a\,x_{2,a}\,\varepsilon_a^{2} \hspace{.5cm}&  1 \\[5pt]  
             \end{array} \right)\,.
\eeq
It can be easily checked that these two matrices diagonalize the Yukawa and respect unitarity up to $\mathcal{O}(\varepsilon_a^{4})$.\\
The CKM is defined in terms of the up and down-quark Left-rotation matrices as
\beq
V_{\rm CKM}\equiv P_{L,u}^{*}\,V_{L,u}^{\dagger}\,V_{L,d}\,P_{L,d} .
\eeq
Given the expansion parameter in the up sector being three times smaller than in the down sector, we can convince ourself that the $V_{L,u}^{\dagger}$ part gives only higher order corrections to the LO structure, so in this qualitative discussion we can consider $V_{\rm CKM}\simeq V_{L,d}$. We can now make use of the re-phasing matrices $P_{L,u}$ and $P_{L,d}$ in order to make the 11, 22, 33, 12, 23 entries real in the CKM matrix. Looking at Eq.(\ref{eq:VLa}) we see that we need to get rid of $\theta_{12}=(\gamma_d -\delta_d), \theta_{23}=\delta_d$, which may can be achieved straightforwardly by
\beq 
  P_{L,d} ~ = ~ P_{L,u}^{*} ~ =\left(\begin{array}{ccc}
              1 \hspace{.5cm}& 0  \hspace{.5cm}& 0 \\[5pt]
              0 \hspace{.5cm}& e^{i\,(\gamma_d -\delta_d)}\hspace{.5cm}& 0\\[5pt]
              0 \hspace{.5cm}& 0 \hspace{.5cm}& e^{i\,\delta_d}  
             \end{array} \right) \,.
\eeq
Additionally, to keep real and positive Yukawa couplings after this rephasing, we must absorb these undesired phases in the right re-phasing matrices, as
\begin{align}
P_{R,e}~&=~Diag\{e^{-i\,(2\gamma_{e}-\delta_{e})},e^{-i\,(\pi+\delta_{e})},1\}\\
P_{R,u}~&=~Diag\{e^{-i\,\pi},e^{-i(\gamma_{d}-\delta_{d})},e^{-i\,\gamma_{d}}\}\\
P_{R,d}~&=~Diag\{e^{-i\,(\pi+2\gamma_{d}-\delta_{d})},e^{-i\,\gamma_{d}},e^{-i\,\gamma_{d}}\}.  
\end{align} 
These same transformations must be performed on the soft-mass matrices. The results may be found in Section \ref{sec:analysis}.

\acknowledgments

We thank M. Jay P\'erez for comments on the manuscript. IdMV thanks the Universitat de Valencia for its hospitality.
IdMV acknowledges
funding from Funda\c{c}\~{a}o para a Ci\^{e}ncia e a Tecnologia (FCT) through the
contract IF/00816/2015, partial support by Funda\c{c}\~ao para a Ci\^encia e a Tecnologia (FCT,
Portugal) through projects CFTP-FCT Unit 777 (UID/FIS/00777/2013) and CERN/FIS-PAR/0004/2017 which are partially funded through POCTI (FEDER), COMPETE, QREN and EU.
AM, ML and OV acknowledge partial support from Spanish MINECO under grant FPA2014-54459-P, MICINN under grant FPA2017-84543-P and by the Severo Ochoa Excellence Program under grant SEV-2014-0398. OV thanks “Generalitat Valenciana” for support under grant PROMETEO2017-033. 
AM acknowledges support from La-Caixa-Severo Ochoa scholarship. All Feynman diagrams have been drawn using Jaxodraw \cite{Binosi:2003yf}.

\end{document}